\documentclass{aa}  

\usepackage{graphicx}
\usepackage{color}
\usepackage[dvipsnames]{xcolor}
\usepackage{subcaption}
\usepackage{multirow}
\usepackage{float}

\usepackage{txfonts}
\usepackage{wasysym}
\newcommand{\Rearth}{R$_{\oplus}$ }
\begin{document}

   \title{Accretion of primordial H-He atmospheres in mini-Neptunes:\\ the importance of envelope enrichment}
   \titlerunning{Accretion of primordial H-He atmospheres in mini-Neptunes}
   \author{M. Mol Lous
          \inst{\ref{UZH}, \ref{UniB}}, 
          C. Mordasini \inst{\ref{UniB}},
         R. Helled 
          \inst{\ref{UZH}}
          }

   \institute{
             Institut f\"ur Astrophysik, Universit\"at Z\"urich, Winterthurerstr. 190, CH-8057 Z\"urich, Switzerland \label{UZH}
             \and 
             Weltraumforschung und Planetologie, Physikalisches Institut, Universit\"at Bern, Gesellschaftsstrasse 6, 3012 Bern, Switzerland \label{UniB}
             \\
             \email{marit@ics.uzh.ch}
             }

   \date{Received 12/20/2023; accepted 02/15/2024}

 
  \abstract
   {
   Out of the more than 5,000 detected exoplanets a considerable number belongs to a category called 'mini-Neptunes'. Interior models of these planets suggest that they have some primordial, H-He dominated atmosphere. As this type of planet does not occur in the solar system, understanding their formation is a key challenge in planet formation theory. Unfortunately, quantifying how much H-He planets have, based on their observed mass and radius, is impossible due to the degeneracy of interior models.
    }
   {Another approach to estimate the range of possible primordial envelope masses is to use formation theory. As different assumptions in planet formation can heavily influence the nebular gas accretion rate of small planets, it is unclear how large the envelope of a protoplanet should be. We explore the effects that different assumptions on planet formation have on the nebular gas accretion rate, particularly by exploring the way in which solid material interacts with the envelope. This allows us to estimate the range of possible post-formation primordial envelopes. Thereby we demonstrate the importance of envelope enrichment on the initial primordial envelope which can be used in evolution models.
   }
   {We apply formation models that include different solid accretion rate prescriptions. Our assumption is that mini-Neptunes form beyond the ice-line and migrate inward after formation, thus we form planets in-situ at 3 and 5 au. We consider that the envelope can be enriched by the accreted solids in the form of water. We study how different assumptions and parameters influence the ratio between the planet's total mass and the fraction of primordial gas.}
   {The primordial envelope fractions for small- and intermediate-mass planets (total mass below 15 M$_{\oplus}$) can range from 0.1\% to 50\%. Envelope enrichment can lead to higher primordial mass fractions. We find that the solid accretion rate timescale has the largest influence on the primordial envelope size.}
   {Primordial gas accretion rates onto small planets can span many orders of magnitude. Planet formation models need to use a self-consistent gas accretion prescription.}


   \maketitle
%

\section{Introduction}
\begin{figure*}
    \centering
    \includegraphics[width=\hsize]{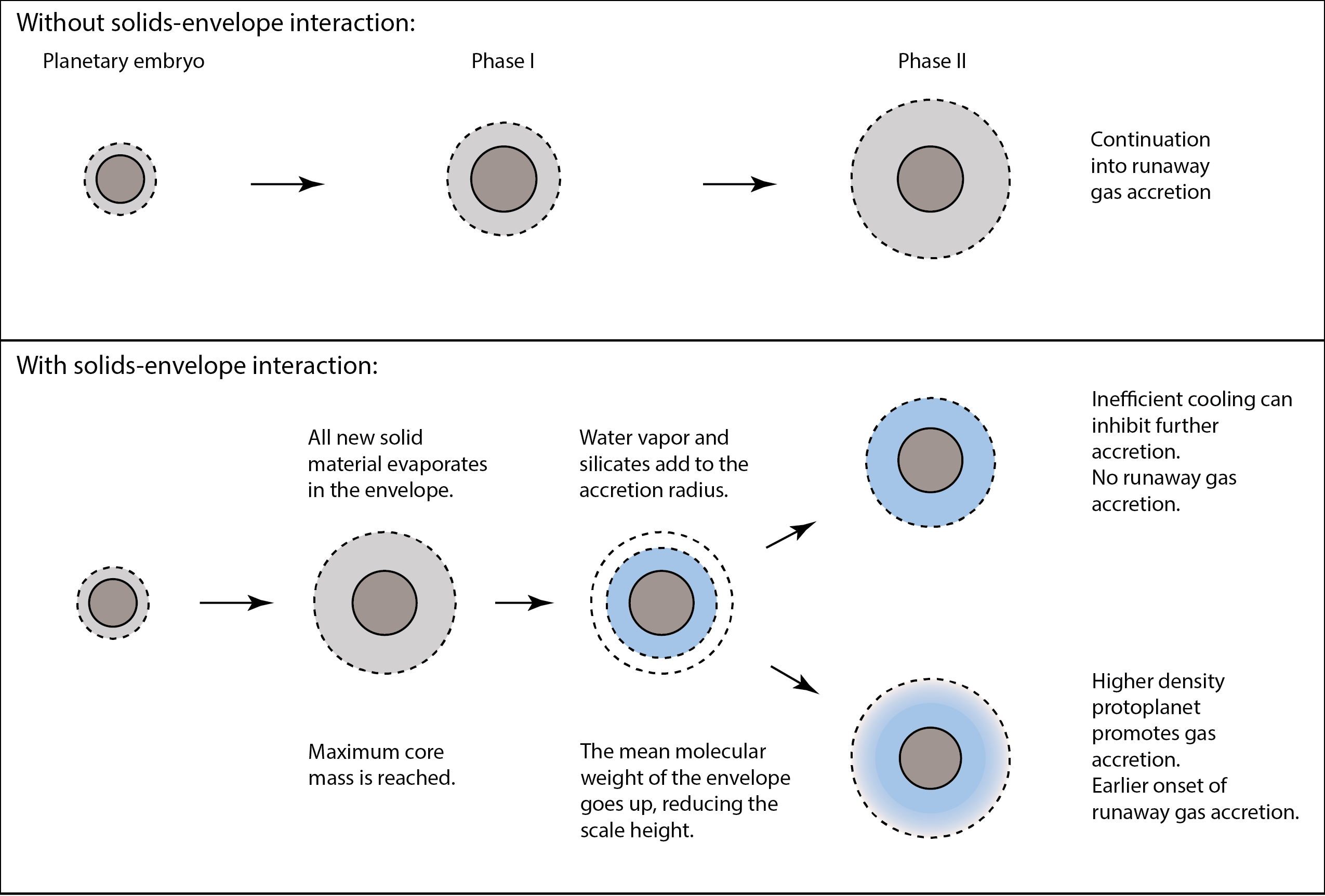}
    \caption{A schematic overview of pre-runaway envelope accretion. When the solid material does not interact with the envelope the gas accretion is initially determined by the size of the core and the strength of the accretion luminosity (Phase I). The core stops growing when there is no more solid material available, so that the envelope accretion rate is determined by the cooling timescale of the protoplanet (Phase II). If Phase II is efficient the planet can reach the critical mass within the lifetime of the protoplanetary disk. It will go into runaway accretion and become a gas giant.
    Gas accretion from the nebula is different when the interaction of the solid material with the envelope is considered. Part of the ice and/or silicates will vaporize rather than reaching the core in solid form. This increases the envelope metallicity. The increased metallicity can on the one hand inhibit further gas accretion by increased opacities in the envelope, which hinders cooling. On the other hand the increased mean molecular weight increases the mean density of the envelope which promotes gas accretion.
    }
    \label{fig:schematic}
\end{figure*}

Currently, more than 5,000 exoplanets have been detected. Many of these planets have sizes larger than Earth but smaller than Neptune \citep{Howard_2012, Fressin_2013, Fulton_2017}, and are commonly referred to as mini-Neptunes. Despite the degeneracy in exoplanetary characterization, interior models indicate that mini-Neptunes consist of non-negligible  hydrogen and helium (H-He) envelopes \citep{Weiss_2014, Rogers_2015, Wolfgang_2015, Jin_2018, Otegi2020, Bean_2021}. These H-He envelopes are thought to be accreted from the protoplanetary disk during the planetary growth. These atmosphere are then  retained despite evolutionary atmosphere loss processes such as photo-evaporation, and therefore can be considered as  {\it primordial envelopes}.
Constraining the initial mass of primordial envelopes of intermediate-mass exoplanets is  a key objective in exoplanet science. For example, constraining the initial mass of the envelopes could provide a solution to the conundrum of the 'radius valley', which is the lack of observed planets with radii between 1.5 \Rearth and 2 \Rearth \citep{Fulton_2017}. In addition, planets with primordial envelopes could be habitable \citep[e.g.][]{Madhusudhan_2021, Mol_Lous_2022}, but one of the major concerns is that a planet must accrete a specific amount of a primordial envelope.
\\ 
Calculating the primordial envelope mass for a given exoplanet is extremely challenging. The prevailing exoplanet measuring techniques only yield radii and masses, through transit measurements and radial velocity detection, respectively. Solving the interior composition of a planet knowing only the mean density and irradiation temperature is a highly degenerate problem \citep{Dorn_2015, Shah_2021, Haldemann_2023}. Additionally there are large errors in the measurements of radii and masses of exoplanets as these are derived in relation to stellar radii and masses, of which the values are not always well constrained \citep{Otegi_2020}.
\\
It is likewise difficult to constrain the size of primordial envelopes from planet formation models. The standard model for planet formation is core accretion \citep{Mizuno_1980, Pollack_1996, Alibert_2005, Helled_2014}. In this scenario, planet formation begins with a solid (heavy-element) core and once this reaches $\sim$0.1 M$_{\oplus}$ the planet starts to accrete a gaseous envelope. Often, planet formation models  predict larger (i.e., more massive) envelopes than the ones inferred for the observed planetary population  \citep[e.g.,][]{Rogers_2021}. There are several possible explanations, including the large uncertainty in the opacities of planetary envelopes \citep{Ormel_2014, Mordasini_2014}, underestimating the role of collisions in atmosphere removal \citep{Denman_2020} or a boil-off phase during disk dispersal \citep{Rogers_2023}. Interestingly, three-dimensional models which include gas-exchange with the surrounding disk predict smaller accreted envelopes than one-dimensional models at an orbital distance of 0.1 AU around a sun-like star \citep{Ormel_2015, Cimerman_2017, Moldenhauer_2021} and it is still unknown whether this inefficiency remains significant at further radial distances such as 3 or 5 au. 
\\
Another physical mechanisms that can greatly alter the accretion of primordial gas in (1D) formation models are the solid-envelope interactions in the planetary envelope during the planetary growth. 
As they grow, protoplanets can accrete solid material and gas simultaneously. Solid material, in the form of planetesimals or pebbles, travels through the envelope and can fragment or ablate. This can enrich the envelope in heavy elements instead of simply being added to the core \citep{Pollack_1986, Podolak_1988}. This process is sometimes referred to as envelope pollution. 
This heavy-element enrichment can have two competing consequences on the planetary growth timescale and the planetary composition. On the one hand, enrichment can increase the envelope's opacity, and therefore delay the planetary contraction and inhibiting the further accretion of nebular gas. 
On the other hand, heavy-element enrichment increases the mean molecular weight of the envelope, which enhances the gas accretion rate. A schematic overview of envelope accretion with and without the consideration of solid-envelope interactions is shown in Figure \ref{fig:schematic}.
\\
Many previous studies have already demonstrated that envelope enrichment plays an important role in planet formation.
Specifically pebbles are quick to ablate and fragment \citep{Ormel_2010, Lambrechts_2014b, Alibert_2017, Chambers_2017, Brouwers_2018, Valletta_2019, Brouwers_2020}, but planetesimals have been shown to interact with the envelope and alter the formation process as well \citep{Stevenson_1982, Hori_2011, Pinhas_2016}. 
Estimates of the maximum core mass that a planet can grow range from $\sim$ 0.1 M$_{\oplus}$ to $\sim$ 5 M$_{\oplus}$ \citep{Pollack_1986, Mordasini_2006, Lozovsky_2017, Alibert_2017, Brouwers_2018, Steinmeyer_2023}. This range is a result of the  different assumptions on planetesimal or pebble sizes, composition, and material strength.
\\
\citet{Valletta_2020} simulated the formation of Jupiter and Saturn, accounting for envelope enrichment where the heavy elements were represented by water. It was found that including envelope enrichment in a self-consistent way (equation of state and opacity calculation) decreases the growth timescale of Jupiter and Saturn. This result is in line with previous studies focusing on giant planet formation \citep{Stevenson_1982, Hori_2011, Venturini_2015, Venturini_2016, Venturini_2017}, but it is not entirely clear if this can be accepted as a general result. It remains a possibility that in some cases the planet cannot cool efficiently enough to trigger runaway gas accretion (see Figure \ref{fig:schematic}). For example, \citet{Wu_2023} showed that if icy pebbles sublimate outside of the accretion radius and enrich the local gas, this decreases the nebular gas accretion efficiency. Furthermore assumptions on mixing efficiency, the composition of the accreted solid material and the strength of the grain opacity can steer the outcome of a one-dimensional planet formation simulation.\\
The objective of this work is to investigate how the accretion rates of gas depend on the model assumptions when envelope enrichment is considered. We follow  \citet{Valletta_2020} and employ a 1-dimensional planet formation model that considers the ablation and fragmentation of the solid material (represented by water ice). 
We focus on the investigation of the envelope's composition of the forming planets before they reach runaway gas accretion. We consider various formation locations, protoplanetary disk properties, and solid accretion rates. 
We also investigate the formation timescales to assess whether the planet is expected to reach the runaway gas accretion and become a gas giant planet. \\
Our paper is organized as follows. In Section \ref{sec:methods} we present our model setup. In Section \ref{sec:results} we present our results for the gas accretion rates for different planets. We distinguish between gas accretion with and without the enrichment of solid materials. The distribution of possible H-He envelope masses within the explored parameter space is given. In this section we also demonstrate the importance of basic assumptions, such as the mixing of supercritical water with H-He, on our results. In Section \ref{sec:discussion} we further test assumptions on mixing and the smoothing of the deposition profile. In this section we also address the likelihood that planets form with the required amount of H-He to allow for surface liquid water. The limitations to our model are discussed in Section \ref{sec:caveats}. Finally in Section \ref{sec:concl} we summarize our findings.
\section{Methods} \label{sec:methods}
The formation simulations are based on a modified version of the MESA code\footnote{Version 10108 of MESA is used and compiled by the MESA SDK that was released simultaneously.} \citep{Paxton_2011, Paxton_2013, Paxton_2015, Paxton_2018, Paxton_2019} which was properly adapted to simulate planet formation and evolution \citep{Valletta_2020,Muller_2020a,Muller_2020b}. The formation model is similar to the one used in \citet{Valletta_2020}, with some modifications as discussed below. \\ 
The initial model has a core mass of 0.1 M$_{\oplus}$ and an envelope of 10$^{-6}$ M$_{\oplus}$. The initial envelope metallicity is 0.03, but drops to zero at the beginning of the evolution when pure H-He is accreted and envelope enrichment is not yet significant. Three different solid accretion rate prescriptions are considered to compute the solid accretion rate $\dot{M_{\text{Z}}}$. Based on planetesimal accretion we use rapid growth \citep{Pollack_1996} and oligarchic growth \citep{Fortier_2013}. We also simulate pebble accretion \citep{Lambrechts_2014}. A summary of these accretion rates are given in Appendix \ref{ap:solid_rates}. For planetesimals we assume a radius of 100 km and for the pebbles one of 10 cm.\\
In the case of planetesimal accretion the simulation starts at 10 kyr. In the pebble accretion case we also assume that the solid accretion rate starts at 10 kyr, but with a smaller initial model, namely 0.01 M$_{\oplus}$. Integrating the solid accretion rate of pebbles (see Appendix \ref{app:pebbles}) from a mass of 0.01 M$_{\oplus}$ at 10 kyr to 0.1 M$_{\oplus}$ gives us the starting times of our planetary embryo. These starting times ($t_{0, \textrm{ peb}}$) depend on the disk conditions and are given in Table \ref{tab:disk}.\\
In the nominal case we set the lifetime of the disk to 10 Myr. As solar-like stars should form within 5 - 10 Myr, this is a long but not unlikely formation time \citep{Pfalzner_2022}. We also consider shorter formation in 3 Myr. We stop our simulations before runaway gas accretion starts, namely when the crossover mass is reached, where the envelope and core are of equal mass \citep{Bodenheimer_1986, Pollack_1996}. 

\subsection{Boundary conditions and disk assumptions}
The outer boundary conditions of our model planetary envelope ($P_{\text{out}}$ and $T_{\text{out}}$) are set equal to the pressure and temperature in the disk. Following \citet{Piso_2014} these are given by scaling relations in distance:
\\
\begin{align}
    P_{\text{out}} = 0.27 \left ( \frac{a}{5.2 \text{ AU}}\right )^{-45 / 14} \text{ Ba},
\end{align}
\begin{align}
    T_{\text{out}} =  150 \left ( \frac{a}{5.2 \text{ AU}} \right )^{-3 / 7} \text{ K},
\end{align}
\\
where $a$ is the orbital distance of the protoplanet. The normalization factors that we apply are higher than the fudicial MMSN (which would be 0.0085 and 60 for pressure and temperature respectively). While this is a significant increase, we find that it does not influence the envelope masses when they are above $\sim$ 0.01 M$_{\oplus}$ and saves computation time. These boundary conditions do play a significant role in the early stages of the protoplanet and could in theory alter the formation path, for example through the onset of fragmentation. Similarly, $P_{\text{out}}$ and $T_{\text{out}}$ are assumed to remain constant in time in this work. More accurate gas accretion simulations would thus require an improved disk model, especially for the cases presented with envelope masses below 0.01 M$_{\oplus}$ after formation. In this work, however, these simplification suffice to demonstrate the importance of envelope enrichment on gas accretion. \\
The planetesimal accretion rates scale linearly with the solid surface density, $\Sigma_{\text{Z}}$, at the location of formation (see Appendix \ref{app:rapid} and \ref{app:olig}). The initial solid surface density $\Sigma_{\text{Z, 0}}$ is given by:
\\
\begin{align} \label{eq:Sigma_init}
    \Sigma_{\text{Z, 0}} =  C_{1} \left ( \frac{a}{5.2 \text{ AU}} \right )^{-1}  \text{ g cm}^{-2},
\end{align}
\\
The solid surface density decreases as solids gets accreted onto the planet.\\
The pebble accretion rate has a linear dependency on the gas surface density at 1 au, $\beta$ (see Appendix \ref{app:pebbles}). $\beta$ is set by an initial gas surface density $\beta_{0}$ and decreases exponentially in time ($t$):
\\
\begin{align} \label{eq:beta}
    \beta = \beta_{0} \exp(-t / \tau_{\textrm{disk}}), 
\end{align}
\\
where $\tau_{\textrm{disk}}$ is the gas disk lifetime which we fix to 3 Myr. 
$C_{1}$ and $\beta_{0}$ are used as free parameters to account for disks of different masses. We set $C_{1}$ to 5, 7.5 or 10 g cm$^{-2}$ for a \textit{light}, \textit{medium} or \textit{heavy} disk, respectively. Values of $\beta_{0}$ for these three disk types are set to 250, 500 or 750 g cm$^{-2}$. The corresponding values of $\Sigma_{Z, 0}$ and $\Sigma_{g, 0}$ at 3 or 5 au are listed in Table \ref{tab:disk}.\\
The inner boundary of the envelope model is the core. The luminosity at the core-envelope interface is determined by the accretion luminosity:
\\
\begin{equation}
    L_{\text{acc}} = \frac{G M_{\text{c}} \dot{M_{\text{z}}}}{R_{\text{c}}} \left ( 1 - f_{\textrm{abl}}\right ),
\end{equation}
\\
where $G$ is the gravitational constant and $f_{\textrm{abl}}$ is the fraction of solid material which is ablated or fragmented in the envelope. $M_{\text{c}}$ and $R_{\text{c}}$ are the core mass and radius respectively, where the value of $R_{\text{c}}$ is calculated assuming a constant core density of 3.2 g cm$^{-3}$, regardless of the composition of the accreted material. The significance of this simplification is considered in Appendix \ref{subsec:MR}. 
\begin{table*}[]
\centering
\begin{tabular}{|ll|l|ll|}
\hline
 &  & \begin{tabular}[c]{@{}l@{}}Rapid accretion\\ Oligarchic accretion\end{tabular} & \multicolumn{2}{l|}{Pebble accretion} \\ \cline{3-5} 
 &  & $\Sigma_{\text{Z, 0}}$ (g / cm$^{-2}$) & \multicolumn{1}{l|}{$\Sigma_{\text{g, 0}}$ (g cm$^{-2}$)} & $t_{0, \textrm{ peb}}$ (kyr) \\ \hline
\multicolumn{1}{|l|}{\textit{Heavy disk}} & 3 au & 17.33 & \multicolumn{1}{l|}{250} & 34.4 \\ \cline{2-5} 
\multicolumn{1}{|l|}{} & 5 au & 10.4 & \multicolumn{1}{l|}{150} & 38.7 \\ \hline
\multicolumn{1}{|l|}{\textit{Medium disk}} & 3 au & 13 & \multicolumn{1}{l|}{167} & 49.2 \\ \cline{2-5} 
\multicolumn{1}{|l|}{} & 5 au & 7.8 & \multicolumn{1}{l|}{100} & 56.5 \\ \hline
\multicolumn{1}{|l|}{\textit{Light disk}} & 3 au & 8.67 & \multicolumn{1}{l|}{83} & 101 \\ \cline{2-5} 
\multicolumn{1}{|l|}{} & 5 au & 5.2 & \multicolumn{1}{l|}{50} & 119 \\ \hline
\end{tabular}\caption{The assumed initial solid surface density and initial gas surface density in the \textit{heavy} disk, \textit{medium} and \textit{light} disk. $\Sigma_{Z, 0}$ is used for the planetesimal accretion, while $\Sigma_{g, 0}$ and t$_{0, \textrm{ peb}}$ are used for the pebble accretion. }\label{tab:disk}
\end{table*}

\subsection{Enrichment from Planetesimals or Pebbles} \label{sec:method_enrich}
\begin{figure}
    \centering
    \includegraphics[width=0.98 \hsize]{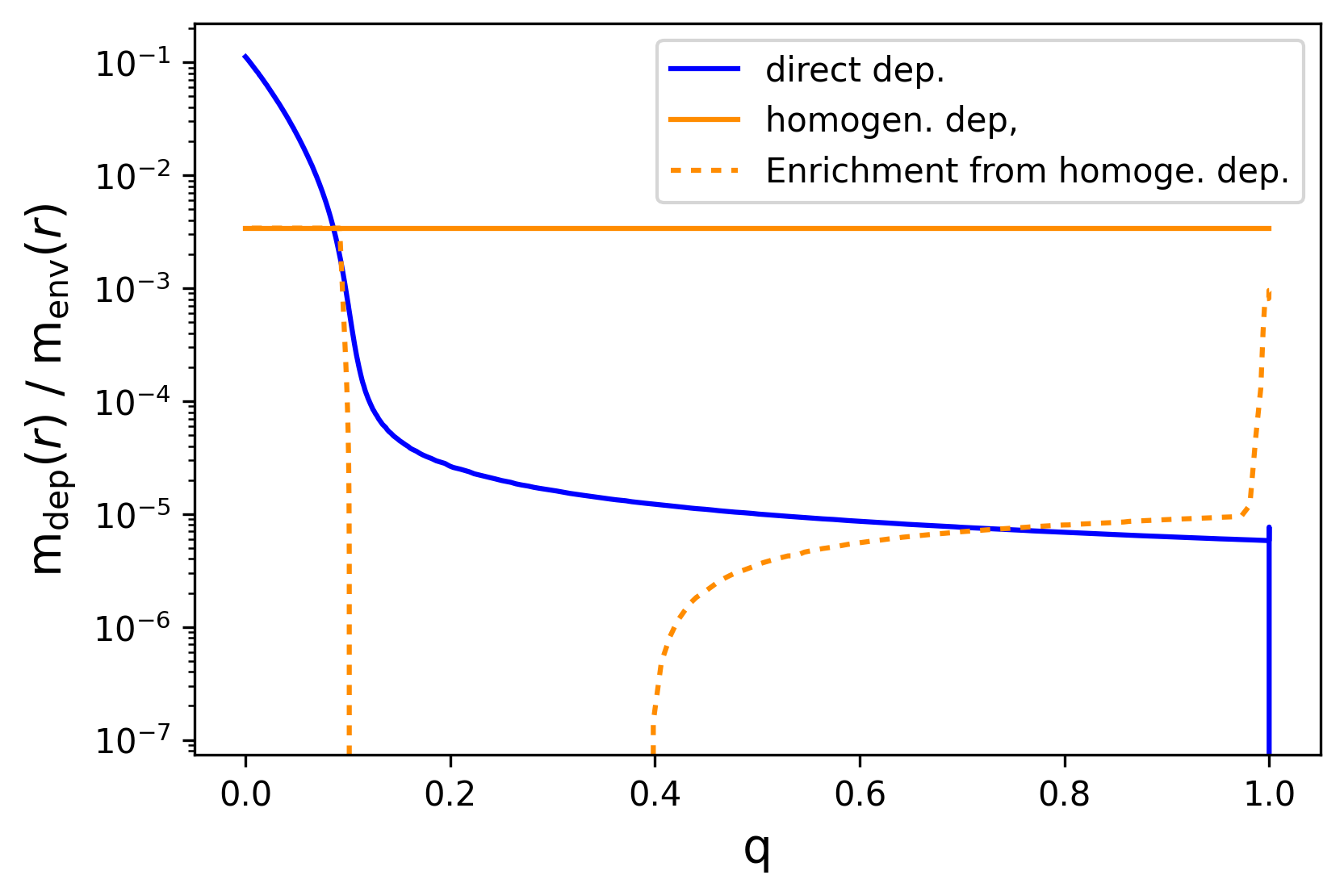}
    \caption{The difference between the two deposit models: \textit{direct deposit} (blue) and \textit{homogeneous deposit} (orange). The x-axis gives the normalized mass coordinate of the envelope $q$. This figure specifically shows the deposit models for Oligarchic growth at 47 kyr when the envelope mass is 1.5 $\times 10^{-5}$ M$_{\oplus}$. There is no fragmentation yet. The ablation enriches the envelope metallicity up to 10\% in the most inner region. The \textit{homogeneous deposit} has the same total deposited mass, but smoothed. However, the actual enrichment is not homogeneous, as shown by the orange dashed line, due to some layers already being saturated. For different formation conditions and at different times the difference between the deposition and the actual enrichment changes.}
    \label{fig:dep_models}
\end{figure}

The interaction between the accreted solids and the envelope is considered via the fragmentation and/or ablation of the solids (planetesimals or pebbles). The calculation of the value of $f_{\textrm{abl}}$ is given in Appendix \ref{app:abl_frac}. This method also gives the deposition profile $m_\textrm{dep}(r)$ at radius $r$. The amount of water vapor added to the envelope is the product of $f_{\textrm{abl}}$ and the solid accretion rate.\\
We consider two deposition methods. The first is \textit{direct deposit}. In this method the mass is deposited at the radial locations where ablation and fragmentation occur. For example: if a planetesimal fragments at radius $r$ and there is no prior ablation, the water mass in layer $m(r)$ is enhanced by the solid accretion rate. This increases the metallicity.\\
The second method, \textit{homogeneous deposit}, is the default in this work. It assumes that the mass deposition is completely smoothed over the envelope which has total mass M$_{\textrm{env}}$. This means that the amount of added heavy material is distributed over all layers, normalized to the layer's mass, where:
\\
\begin{equation}
    m_{\textrm{dep}}(r) = f_{\textrm{abl}} \times \dot{M_{\text{Z}}} \frac{m(r)}{M_{\textrm{env}}}.
\end{equation}
\\
As an illustration, Figure \ref{fig:dep_models} shows the difference between \textit{direct deposit} and \textit{homogeneous deposit} specifically for a planet growing by oligarchic growth at 5 au after 47 kyr. The core mass is still the initial 0.1 M$_{\oplus}$ and the envelope mass is 1.5 $\times 10^{-5}$ M$_{\oplus}$. The envelope is too small to cause fragmentation of the planetesimals and thus there is only ablation. The fraction of ablated material increases towards the interior of the envelope. The \textit{homogeneous deposit} is completely smoothed over all layers, but the total deposited material adds up to the same as for \textit{direct deposit}.\\
While at every timestep the deposition of heavy material is done homogeneously, this does not necessarily mean that the composition in the envelope is homogeneous. This is for two reasons. First of all because there is a gas accretion of pure H-He with zero metallicity added to the outer layers of the envelope. The inner layers, which are older, will have been exposed to envelope enrichment for longer and thus have a higher metallicity. This will create a compositional gradient unless the Ledoux criterium is met, in which case the convective region will become homogeneously mixed. Secondly, we consider a maximum metallicity in each layer and if this is already reached there is no enrichment. An example of this is shown in Figure \ref{fig:dep_models}. The orange solid line shows the deposit profile, while the dashed orange line shows the actual enrichment. The difference is due to some layers in the envelope already being saturated, or close to saturation, so that it is not possible to deposit all the mass without condensation. There are two criteria that could limit the amount of water that can be deposited in a certain layer.\\
First, we check the material state of H$_2$O based on the temperature of the layer and from there calculate the maximum number density of water in layer $r$:
\\
\begin{equation}\label{eq:number_max}
    n_{\text{H}_{2}\text{O, max}} (r) =
\begin{cases}
 & \frac{P_{\text{sat}}(r)}{P(r) + P_{\text{sat}}(r)} \; \; \; \; \; \text{ if } T(r) < \text{T}_{\text{crit}} \\ 
 &  Z_{\textrm{max}} \; \; \; \; \; \; \; \; \; \; \; \; \; \text{if } T(r) > \text{T}_{\text{crit}}
\end{cases}
\end{equation}
\\
$P(r)$ and $T(r)$ are the pressure and temperature at radius $r$. T$_{\text{crit}}$ is the critical temperature of 647.096 K.
In the cases where the temperature is below 647 Kelvin, we apply the vapor-liquid phase boundary from \citet{Wagner_2002} to calculate the saturation pressure of water $P_{\text{sat}}$ as follows: 
\\
\begin{equation}
    \frac{P_{\text{sat}}(r)}{\text{P}_{\text{crit}}} = \exp (\frac{\text{T}_{\text{crit}}}{T(r)} \left(  a_1 \nu + a_2 \nu^{1.5} + a_3 \nu^{3} + a_4 \nu^{3.5} + a_5 \nu^{4} + a_6 \nu^{7.5} \right )).
\end{equation}
\\
Here $\text{P}_{\text{crit}}$ is the critical pressure, 220.64 bar. $\nu = \left ( 1 - \frac{T}{\text{T}_{\text{crit}}}\right )$. The other variables are a$_{1}$ = -7.85951783, a$_{2}$ = 1.84408259, a$_{3}$ = -11.7866497, a$_{4}$ = 22.6807411, a$_{5}$ = -15.9618719, a$_{6}$ = 1.80122502.\\
If the temperature exceeds the supercritical temperature of water we impose a limit to the water enhancement, $Z_{\textrm{max}}$. Supercritical water and H-He are expected to be highly miscible \citep{Soubiran_2015}, suggesting that $Z_{\textrm{max}} = 1$. However in our nominal model we set this value lower to 0.9 to ensure that we do not artificially create a loss of H-He. This would happen if too much H-He inside the envelope is replaced by water without a sufficiently high primordial gas accretion rate. We also want to investigate the significance of this miscibility and additionally consider this maximum metallicity to be to 0.5.\\
The second criterion for water deposition is that the deposited material can only be as massive as the shell in which it is deposited. As MESA uses mass coordinates, this criterion ensures that there is no Rayleigh–Taylor instability created through the deposition. While this an artificial limit, we argue that the deposited material we calculate in a certain layer using our one-dimensional model underestimates the smoothing over different layers. Thus, allowing this smoothing of the water deposition profile should better represent the three-dimensional structure.\\
When it is not possible to deposit part of the heavy material in the envelope, we transfer the leftover water mass to the core. \\
Thus the enrichment is only equal to the initial deposition if a layer is not yet saturated, as is demonstrated in Figure \ref{fig:dep_models}. In this specific case for the inner 10\% of envelope mass ($q < 0.1$) the critical temperature is exceeded, so that the maximum metallicity is much higher than for $q > 0.1$. Furthermore the outer envelope ($q > 0.4$) contains newer gas which has not been exposed to as much enrichment, hence the enrichment increases towards the outside of the envelope. If the envelope is not  convective a compositional gradient can be created.\\
Finally, we define the metallicity of the envelope at location $r$ as:
\\
\begin{equation}
    Z(r) = \frac{m_{\text{H}_{2}\text{O}}(r)}{m(r)}. 
\end{equation}
\\
The change in the envelope's metallicity alters the opacities and the equation-of-state of the envelope. The total envelope's metallicity is referred to as Z$_{\textrm{env}}$ and is defined by the total water mass fraction in the envelope:
\\
\begin{equation}
    Z_{\textrm{env}} = \frac{M_{\textrm{env, H}_{2}\text{O}}}{M_{\textrm{env}}}. 
\end{equation}
\\
The opacities are calculated by adding the molecular opacities from \citet{Freedman_2014} and grain opacities from \citet{Alexander_1994}, following \citep{Valencia_2013}. We use an equation-of-state that mixes water with hydrogen and helium taken from \citet{Muller_2020b} (see their Appendix A for details). \\
Finally the heavy element deposition in layer $r$ has two influence on the energy. First of all accretion luminosity is added by:
\\
\begin{equation}
    l_{\text{acc}}(r) = \frac{G M(r) m_{\textrm{dep}}(r) }{r}
\end{equation}
\\
with $M(r)$ the cumulative mass at radius $r$.\\
Secondly the vaporization of water decreases the energy by \citep{Pollack_1986}:
\\
\begin{equation}
    E_{\text{vap}}(r) = - (c_{\text{p}} \times \Delta T + E_{0}) \times m_{\textrm{dep}}(r)
\end{equation}
\\
where $c_{\text{p}} = 4.2 \times 10^{7}$ (erg g$^{-1}$ K$^{-1}$) is the specific heat of water, $E_{0} = 2.8 \times 10^{10}$ (erg g$^{-1}$) is the latent heat of vaporisation and $\Delta T$ is the change of temperature to reach vaporization, which we set to 373 K assuming that the incoming pebble/planetesimal has temperature 0 K.\\
\begin{figure}
    \centering
    \includegraphics[width=0.8 \hsize]{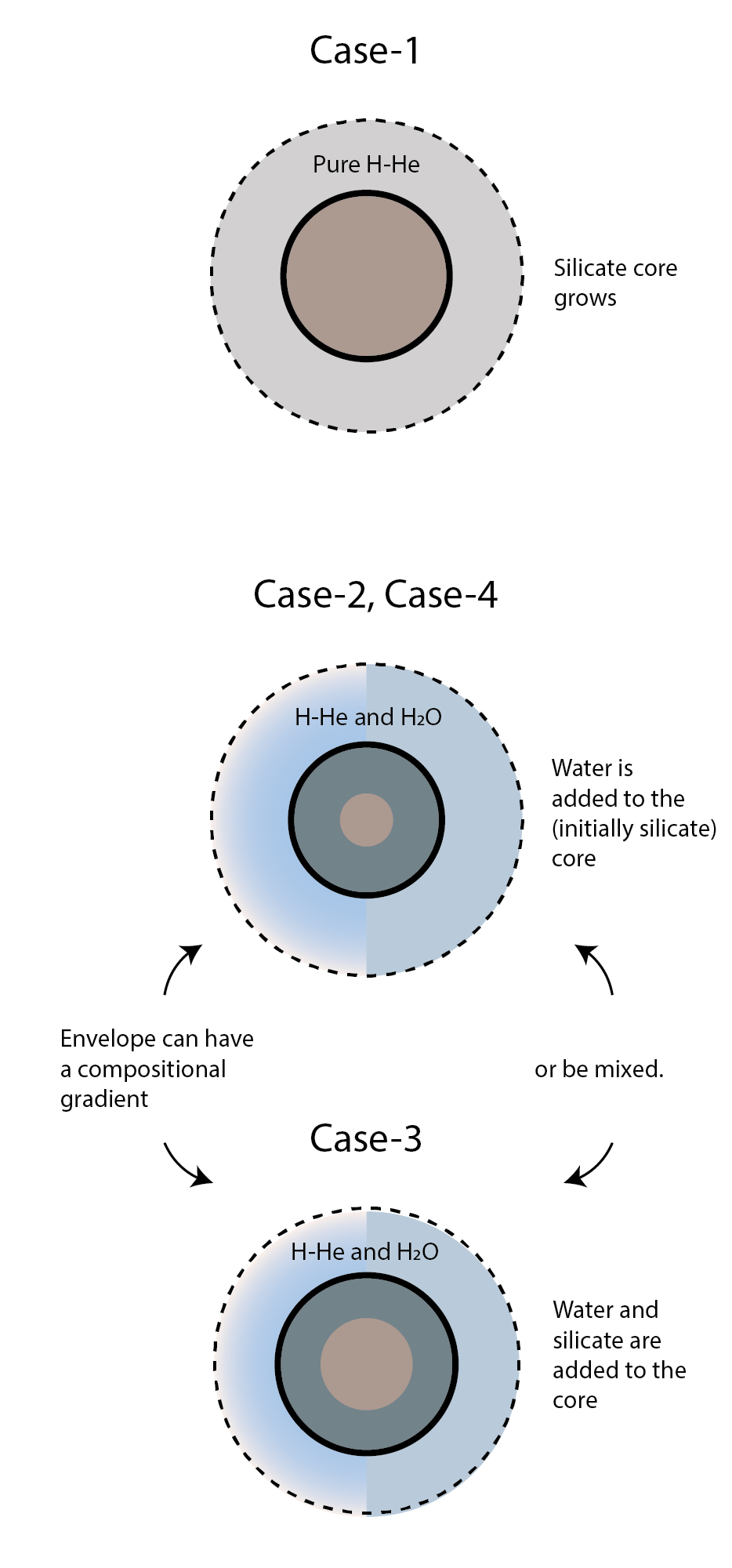}
    \caption{Core and envelope compositions under the different solid-envelope interaction models presented in Table \ref{tab:cases}. The dashed black line represents the outer boundary to the envelope and the solid black line the inner boundary. Everything interior to the black solid line is considered as the core. The composition of the core and whether this is mixed is not considered in this work. Rather a constant core density of 3.2 g / cm$^{3}$ is used. When envelope enrichment is considered this can either create a compositional gradient or there can (a) mixed convective zone(s), as shown by the two halves.
    }
    \label{fig:interior}
\end{figure}
\begin{table}[]
\begin{tabular}{|l|ll|l|l|}
\hline
 & \multicolumn{2}{l|}{\begin{tabular}[c]{@{}l@{}}Ablation \& \\ fragmentation \\ of water\end{tabular}} & \begin{tabular}[c]{@{}l@{}}Solid\\ composition\end{tabular} & \begin{tabular}[c]{@{}l@{}}$Z_{\textrm{max}}$\end{tabular} \\ \hline
Case-1 & \multicolumn{2}{l|}{No} & 100\% silicates & n.a. \\ \hline
Case-2 & \multicolumn{2}{l|}{Yes} & 100\% water & 0.9 \\ \hline
Case-3 & \multicolumn{2}{l|}{Yes} & \begin{tabular}[c]{@{}l@{}}50\% water\\ 50\% silicates\end{tabular} & 0.9 \\ \hline
Case-4 & \multicolumn{2}{l|}{Yes} & 100\% water & 0.5 \\ \hline
\end{tabular} \caption{Solid-envelope interaction models considered in this work. All models assume that silicates do not react with the envelope and directly reach the core.} \label{tab:cases}
\end{table}
\\
In our simulations we distinguish between five types of solid-envelope interactions which are presented in Table \ref{tab:cases}. For simplicity, we neglected the thermal ablation or fragmentation of silicates and focus only on water. Therefore, Case-1 is a reference case without any solid-envelope interactions. All solid material directly reaches the core-envelope boundary and the envelope never increases in metallicity. We consider the other extreme in Case-2. We assume that all the solid material is ice and can enrich the envelope.\\
With Case-1 and Case-2 the most extreme, we use Case-3 and Case-4 to investigate other aspects related to our fragmentation and ablation model. Case-3 is a hybrid between Case-1 and Case-2. Half of the solid material are icy planetesimals/pebbles which can enrich the envelope. The other 50\% of the solid accretion rate consists of rocky material that directly reaches the core and does not interact. Finally, in Case-4 we limit the maximum allowed metallicity $Z_{\textrm{max}}$ (see Equation \ref{eq:number_max}) to 0.5 if the temperature exceeds the critical temperature. Figure \ref{fig:interior} visualizes the effects of these cases on the planet's interior and envelope.
\subsection{Gas accretion}
Gas accretion can occur at every timestep. Following  \citet{Valletta_2019}, gas is added to the planet until the outer radius of the envelope is within a factor 1.1 smaller or larger than the accretion radius. We use the accretion radius as in \citet{Lissauer_2009}. This formulation is based on the common assumption that the planet's  accretion radius must be equal to the smallest of either the Bondi radius or the Hill radius: 
\\
\begin{align}
    R_{\text{acc}} = \frac{GM_{\text{p}}}{c_{\text{s}}^{2} / k_{1} + G M_{\text{p}} / \left ( k_{2} R_{\text{Hill}} \right )}, 
\end{align}
\\
where $M_{\text{p}}$ is the mass of the protoplanet, $c_{\text{s}}$ is the speed of sound at the location of formation, $R_{\text{Hill}}$ is the protoplanet's Hill radius and $k_{1}$ and $k_{2}$ are reduction factors to account for the limited availability of gas at the formation location of the planet. For the small protoplanets considered in this study $k_{1}$ and $k_{2}$ can be set to 1.\\
The first 10 kyr are used to relax the envelope mass. The initial model does not automatically satisfy the criterion that the accretion radius equals the radius of the initial model. How much these two values deviate depends on the orbital distance. We smooth this transition by finding $k_{1}$ and $k_{2}$ values such that the initial model radius is close to the accretion radius. Then we increase $k_{1}$ and $k_{2}$ linearly in time until these are both 1 at 20 kyr.


\section{Results} \label{sec:results}
We perform a grid of simulations with the following variations: the solid accretion rate is rapid, oligarchic or pebbles. The formation location is either 3 or 5 au and the disk is either \textit{heavy}, \textit{medium} or \textit{light} as defined in Table \ref{tab:disk}. An overview of all these results is given in Appendix \ref{app:results}.

\subsection{Solid-envelope interaction affecting and H-He gas accretion} \label{results:examples}

This subsection highlights the effect of all four solid-envelope interaction models on individual formation cases. 
\subsubsection{Rapid Growth}
\begin{figure}
    \centering
    \includegraphics[width=\hsize]{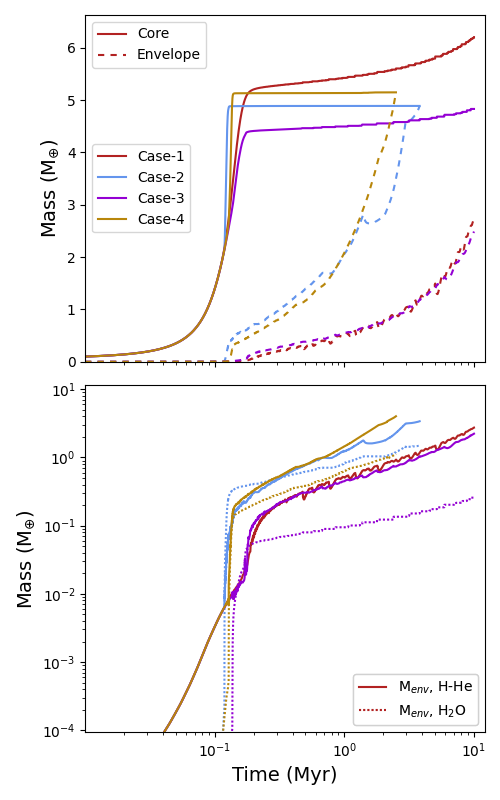}
    \caption{In-situ formation of a planet at 3 au with rapid growth. The simulations are run until the envelope and core are of equal mass or until 10 Myr. The initial solid surface density is 17.33 g cm$^{-2}$ (\textit{heavy} disk). The upper panels shows the mass of the core and the mass of the envelope over time. The lower panel shows the same simulations as in the upper panel, but only the envelope masses over time. Solid lines are the mass from primordial H-He. The dashed lines are the envelope mass from water vapor. In Case-1 there is no water vapor as only hydrogen and helium are accreted from the nebula. 
    }
    \label{fig:growth_rapid}
\end{figure}
\begin{figure}
    \centering
    \includegraphics[width=\hsize]{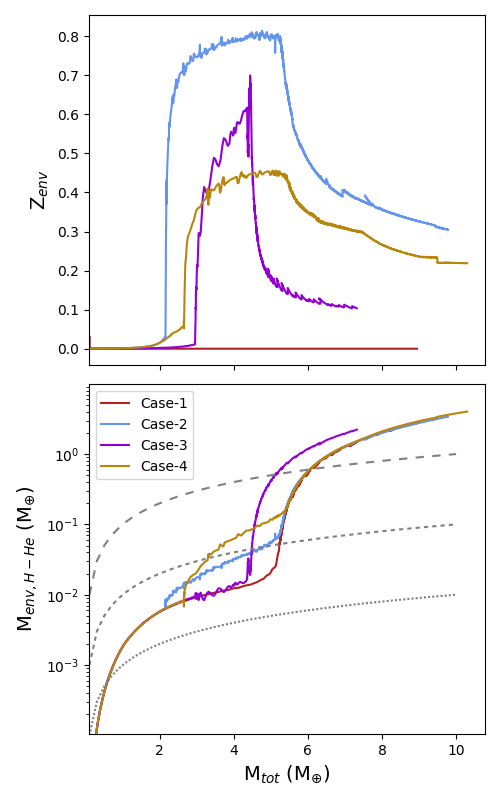}
    \caption{The same simulations presented in Figure \ref{fig:growth_rapid} but showing the heavy element fraction of the envelope (upper panel) and the primordial envelope mass, from pure H-He (lower panel). Both are shown as a function of the total mass (M$_{\textrm{core}}$ + M$_{\textrm{env}}$). The grey dashes lines indicate where primordial envelope mass fractions would be 0.1\%, 1\%, and 10\% of the total mass.}
    \label{fig:growth_2_rapid}
\end{figure}

Figure \ref{fig:growth_rapid} shows the in-situ formation of a planet by rapid growth at 3 au. The initial solid surface density is 17.33 g cm$^{-2}$ (\textit{heavy} disk). The upper panel shows the masses of the core (solid line) and envelope (dashed line) as time  progresses for the various cases. We find that all the cases include both Phase I and Phase II of gas accretion, where the transition between them occurs after $\sim$ 10$^{5}$ yrs at a core mass between 4.4 - 5.2 M$_{\oplus}$. For Case-1 and Case-3 there is still a small increase in core mass during Phase II of gas accretion. At this stage the planet grows through envelope accretion which extends the planetary feeding zone and provides more solid material that can  be accreted by the growing planet. In Case-2 and Case-4 on the other hand, the maximum core mass is reached, as any newly accreted planetesimals fragment and only add water vapor to the envelope. Another distinction is that Case-2 and Case-4 reach a crossover mass after 3.81 Myr and 2.5 Myr, respectively, while Case-1 and Case-3 do not reach crossover mass within 10 Myr. This is because in the former two cases the ablation-fragmentation transition occurs before the feeding zone is depleted and solid accretion is high. This promotes the gas accretion for several reasons. First, the total accretion luminosity is reduced, as a large fraction of the mass is deposited at larger radii and meanwhile the evaporation of water decreases energy locally. Second, the mean molecular weight of the envelope increases so that a more massive envelope can be bound. Similar to previous work we find that the increased opacities due to an increased envelope metallicity do not counteract the mechanisms promoting gas accretion. As such envelope enrichment promotes total envelope accretion.\\
The lower panel of Figure \ref{fig:growth_rapid} shows the envelope's growth, where the contributions of H-He are separated from the water vapor. Since Case-3 has a low-metallicity envelope, the total H-He mass in the envelope is similar to that of Case-1. Note that there would be a larger difference between Case-1 and Case-3 if fragmentation occurred before the solid accretion rate decreases. Case-2 and Case-4 have significantly more massive H-He envelopes at a given time. After $\sim$ 10$^{5}$ years it remains a factor 3 higher than Case-1 and Case-3. We find that  for a short time the water mass  in the envelope exceeds the H-He mass. This occurs during the transition between Phase I and Phase II. However, since subsequently mostly H-He is accreted, the envelopes final atmospheric composition  is dominated by   H-He.\\
During Phase II accretion we find that small amounts of envelope mass can be lost. For Case-1 and Case-3 this concerns small oscillations in the envelope mass which are a result of an oscillating solid accretion rate. These are in turn due to the changes in capture radius, which depends on the internal structure of the envelope (see Appendix \ref{app:rapid}). In other words, when the gas accretion rate is large, the capture radius also increases, promoting a higher solid accretion rate. However, the increase in luminosity from the gas accretion and the solid accretion expand the envelope and increase the radius beyond the accretion radius, which leads to mass loss. While the solid accretion rate remains small (between 10$^{-8}$ M$_{\oplus}$ yr$^{-1}$ and 0) this is sufficient to influence the envelope. Nonetheless, we do find that smoothing the change in capture radius during Phase II gas accretion (by only allowing it to change with 0.1\% every timestep) eliminates these oscillations without altering the final outcome. In Case-2 we find a single instance of mass loss at 1.42 Myr. Similarly to Case-1 and Case-3 this mass loss proceeds from an increase in the capture radius. However, in this case the increased capture radius is due to a change in the internal structure of the envelope, as the size of the convective zone increases.\\
The top panel of Figure \ref{fig:growth_2_rapid} shows the envelope's metallicity as a function of the total planetary mass. For all cases the envelope metallicity peaks when the feeding zone is depleted. The maximum envelope metallicity in Case-2 and Case-3 peaks at  $\sim$0.8. This is expected from the maximum metallicity in supercritical states, $Z_{\textrm{max}}$, set to 0.9. Colder outer layers where water can condense have even lower metallicities which decreases the total envelope metallicity from $Z_{\textrm{max}}$.\\
The lower panel of Figure \ref{fig:growth_2_rapid} shows H-He envelope mass as a function of the total planetary mass. Grey dashed lines give the reference fractions of $f_{\textrm{H-He}}$= 0.1\%, 1\% and 10\%, where $f_{\textrm{H-He}}$ = $M_{\textrm{env, H-He}}$ / $M_{\textrm{p}}$. When the planet is smaller than $\sim$ 2 M$_{\oplus}$ the primordial envelope masses of all cases are similar. At higher masses the solid accretion rate increases and fragmentation occurs, so that the envelope has a significant amount of water vapor which influences  the H-He accretion. At a total mass of 5 M$_{\oplus}$, Case-2 has a factor 2 higher M$_{\textrm{env, H-He}}$ than Case-1 and Case-4 has a factor 5 higher than Case-1. However, at masses above 6 M$_{\oplus}$ the primordial envelope mass of Case-1, Case-2 and Case-4 converge, as Phase II of gas accretion sets in.\\ Interestingly, Case-3 has the transition into Phase II of gas accretion for a lower mass than the other three cases. Compared to Case-1, Case-3 has a lower core mass, as there is always a fraction between 0 and 0.5 of solid material evaporating in the envelope. Also compared to Case-2 and Case-4 the maximum core mass is smaller. This is because Case-2 and Case-4 are more efficient at enhancing the envelope and they reach a stage where $f_{\textrm{abl}}$ equals 1 before the feeding zone is depleted. As a result, envelope accretion accelerates and in this larger envelope planetesimals are captured more easily (i.e. the capture radius as defined in Appendix \ref{app:rapid} increases). Thus, as the solid accretion rate increases, the envelope becomes saturated with water vapor which then allows the core to grow more rapidly as well. This acceleration of core and envelope formation is not evoked in Case-3 because of the later onset of fragmentation.\\

\subsubsection{Oligarchic Growth}
\begin{figure}
    \centering
    \includegraphics[width=\hsize]{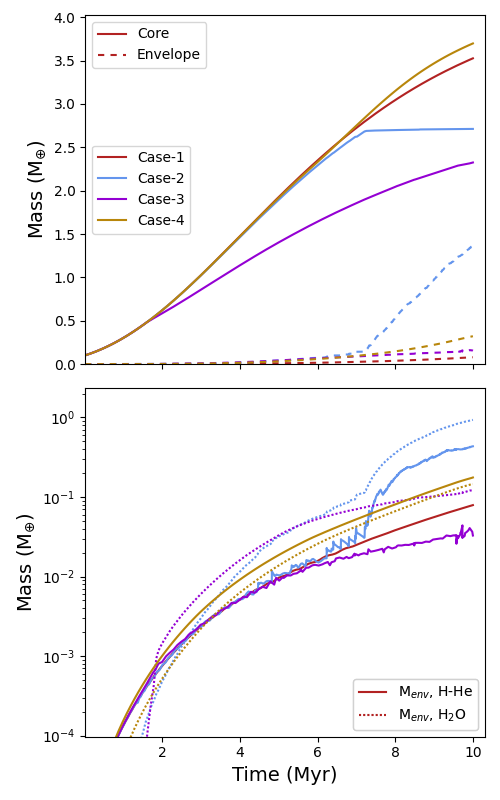}
    \caption{Same as Figure \ref{fig:growth_rapid} but for a planet forming in-situ at 3 AU by oligarchic growth in a \textit{heavy disk}. 
    }
    \label{fig:growth}
\end{figure}
\begin{figure}
    \centering
    \includegraphics[width=\hsize]{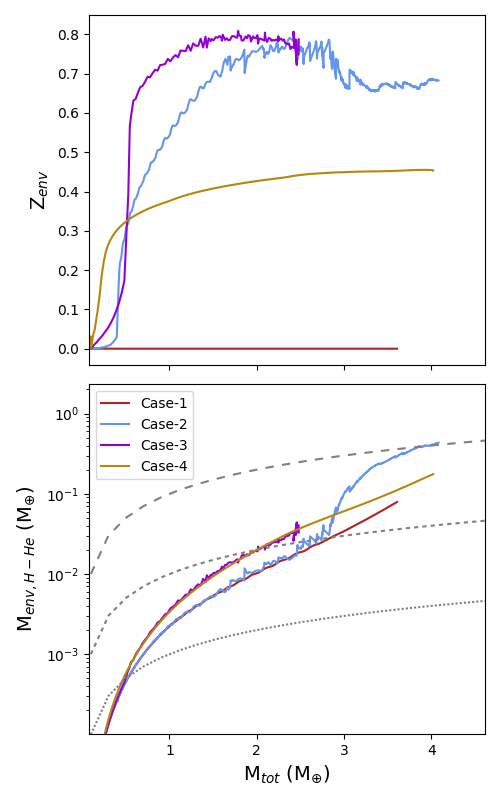}
    \caption{Same as Figure \ref{fig:growth_2_rapid} but for a planet forming in-situ growth at 3 AU by oligarchic growth in a \textit{heavy disk}.}
    \label{fig:growth_1}
\end{figure}

Figures \ref{fig:growth} and \ref{fig:growth_1} demonstrate the effect of envelope enrichment on the  planetary mass and bulk composition as well  as the  formation timescale for oligarchic growth at 3 au. Similar to the previously presented rapid growth, the initial solid surface density is 17.33 g cm$^{-2}$, corresponding to a \textit{heavy disk}.\\
The upper panel in Figure \ref{fig:growth} shows that in Case-2 the core reaches a maximum of 2.7 M$_{\oplus}$. This is notably lower than the core of the planet formed by rapid planetesimal accretion at the same location in a \textit{heavy} disk. The reason for this difference is that the rapid formation has a high solid accretion rate with a large accretion luminosity. This makes the total envelope mass smaller for a given core mass, so that complete fragmentation is reached for a higher core mass in rapid growth. However, the core mass presented in these results also contain evaporated water that could not be held in the saturated envelope layers. Further discussion on the impact of our core model assumptions on the accretion of H-He is presented in Section \ref{subsec:MR}. Case-4 has a larger core accretion rate than any of the other models in the last 3 Myr. This is because Case-4 has a more massive envelope and thus a larger capture radius. Furthermore, because Z$_{\textrm{max}}$ in Case-4 is lower than those in Case-2 and Case-3, the envelope becomes saturated earlier.\\
Since oligarchic growth is much slower compared to rapid growth, it takes longer before the core is massive enough to accrete an envelope with which the solids will interact.
As shown in the lower panel of Figure \ref{fig:growth}, the envelopes in Case-2 and Case-3 become water dominated after 2.7 and 1.8 Myr, respectively, and this compostion persists during the remaining planetary growth. The H-He mass is unchanged for Case-1, Case-2, Case-3 until 7 Myr, while Case-4 always has more H-He.\\
The upper panel in Figure \ref{fig:growth_1} confirms that the envelope metallicity increases at a smaller core mass for oligarchic growth compared to rapid growth. The maximum metallicities also occur at smaller core masses than for rapid growth because there is not enough time to grow larger cores. The lower panel shows that Case-1 and Case-2 have very similar H-He envelope fractions until Case-2 reaches its maximum core mass. Also Case-3 and Case-4 have similar H-He envelope mass fractions. 
Overall, we find that the H-He mass fractions are larger in oligarchic growth than in rapid growth, since the envelopes for a given core mass are larger due to the slower formation. 
\subsubsection{Pebbles} \label{results:pebbles}
\begin{figure}
    \centering
    \includegraphics[width=\hsize]{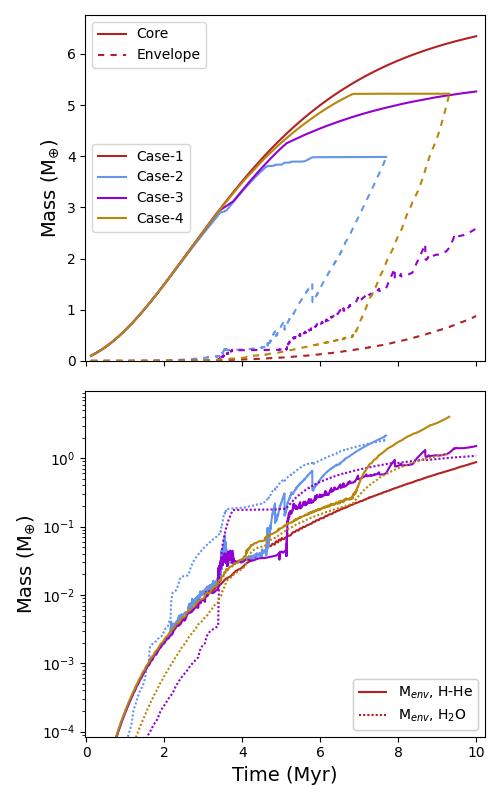}
    \caption{In-situ formation of a planet at 5 au by pebbles. The initial disk conditions are \textit{light}. The upper left panel shows how the core mass and envelope grow in time. The lower left panel shows the envelope mass, with the separated mass contributions of H-He (solid lines) and H$_{2}$O (dashed lines).}
    \label{fig:growth_peb}
\end{figure}
\begin{figure}
    \centering
    \includegraphics[width=\hsize]{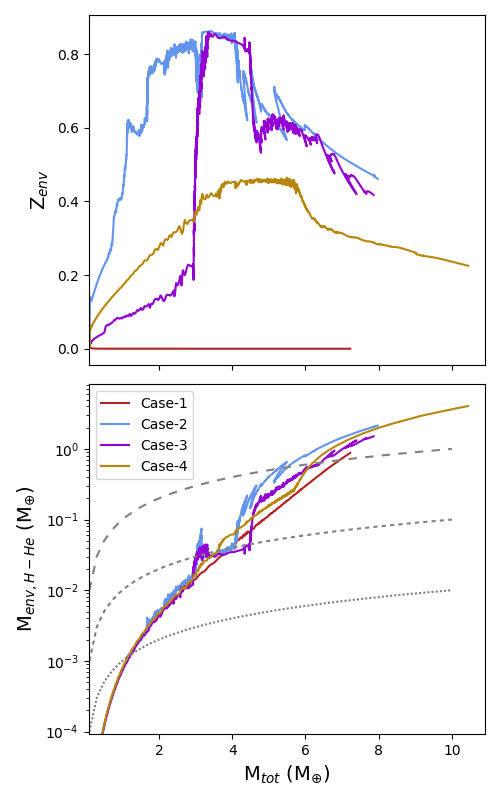}
    \caption{Same simulations as shown in Figure \ref{fig:growth_peb}. Dashed grey lines show H-He envelope mass fractions of 0.1, 0.01 and 0.001.
    }
    \label{fig:growth_peb_2}
\end{figure}

Since pebble accretion is more efficient than planetesimal accretion, we find that most of our pebble simulations reach crossover mass before 3 Myr even when using a later starting time than for the planetesimal accretion. Only in the case of a \textit{light} disk at 5 au we find planets in a pre-runaway state after 10 Myr. As a result, this is the formation scenario we highlight in Figures \ref{fig:growth_peb} and \ref{fig:growth_peb_2}. Due to the small size of pebbles, the value of {\it $f_{\textrm{abl}}$} reaches 1 already at the beginning of the simulation. We use the first 3 kyr of the simulation to smooth {\it $f_{\textrm{abl}}$} from 0 to 1 linearly in time.\\
Figure \ref{fig:growth_peb} shows that Case-1 and Case-3 do not reach a crossover mass while Case-2 and Case-3 reach it after 7.7 and 9.3 Myr, respectively. In Case-2 the maximum core mass is 3.9 M$_{\oplus}$ and for Case-4 it is 5.2 M$_{\oplus}$. Case-1 leads to a core mass of 6.3 M$_{\oplus}$ and an envelope of 0.88 M$_{\oplus}$ after 10 Myr while Case-3 ends with a 5.2 M$_{\oplus}$ core and an envelope of 2.6 M$_{\oplus}$. \\
There are instances of mass loss in Case-2 and Case-3. Contrary to the rapid cases, this is not linked to the coupling between the solid accretion rate and the envelope structure. Instead, this is due to the small size of the pebbles and their immediate fragmentation. In combination with our model set-up, which allows the envelope to be considered 'full' and adds additional water to the core, this can cause the value of $f_{\textrm{abl}}$ to drop when the metallicity is close to saturation. This allows temporary oscillations in the accretion luminosity and possibly, mass loss. These changes in $f_{\textrm{abl}}$ are unphysical and should be modeled more self-consistently in future work. It must be noted, however, that the interaction between icy pebbles and nebular gas can already enhance metallicities at distances further away from the protoplanet than where the gas is bound. In Section \ref{caveats:one_d} we discuss this point and argue that this interaction needs to be well understood before it can be incorporated in one-dimensional models.\\
The envelope's metallicity and primordial envelope mass for the pebble cases are shown in Figure \ref{fig:growth_peb_2}. The metallicities peak at masses of 2 - 6 M$_{\oplus}$. We find that all the enrichment models follow roughly the same relation between H-He mass fraction and total mass, as shown in the lower figure. In comparison to the planetesimal accretion models in Figures \ref{fig:growth_2_rapid} and \ref{fig:growth_1} however these are less smooth. This is because the instant ablation of the pebbles. 

\subsection{H-He mass fractions after formation}
\begin{figure*}
    \centering
    \includegraphics[width=0.95\hsize]{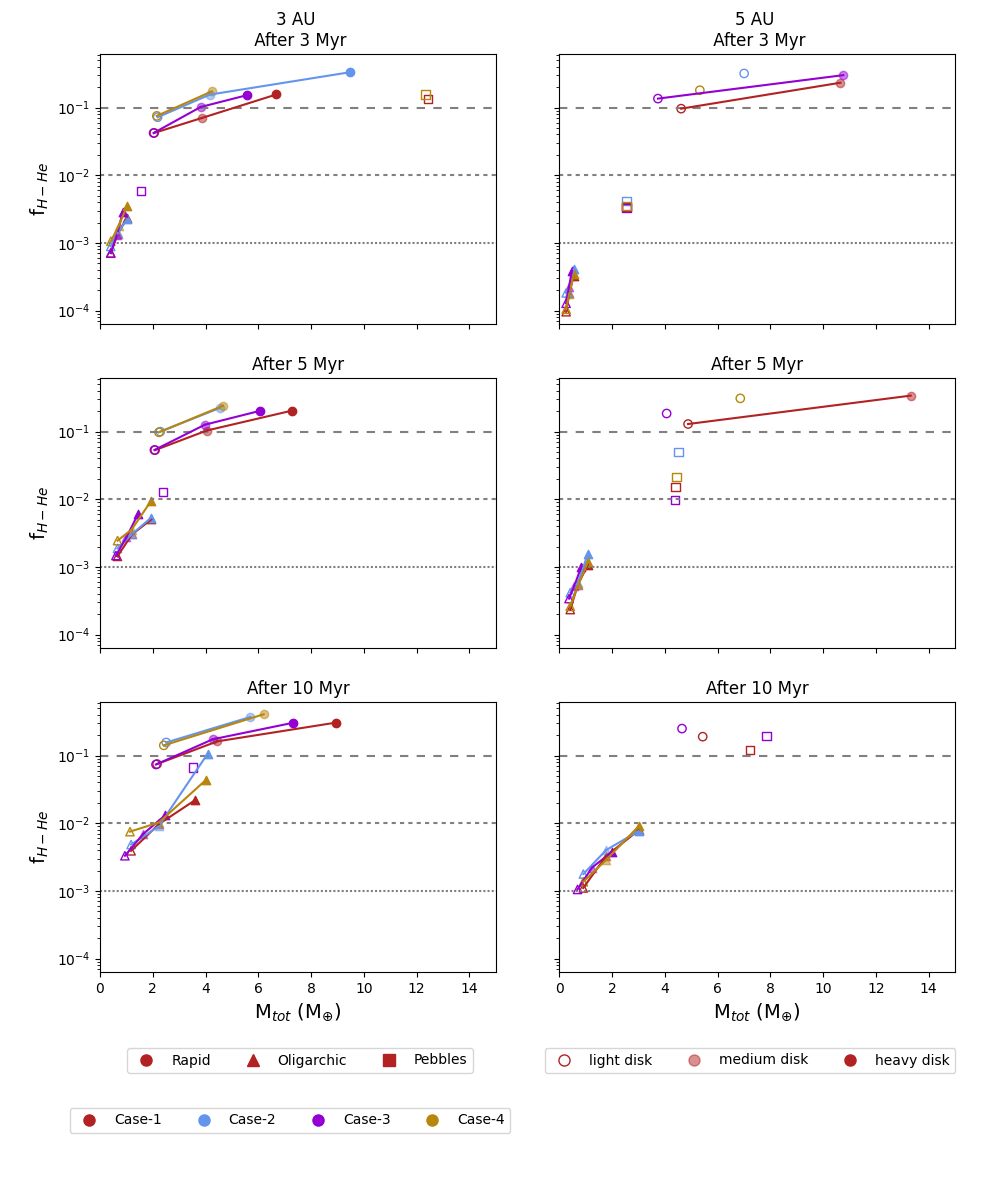}
    \caption{H-He mass fraction after 3 Myr (upper panels), 5 Myr (middle panels), and 10 Myr (lower panels). The left panels show in-situ formation at 3 au and the right panels at 5 au. The colors indicate the heavy-element interaction models. Case-1, Case-2, Case-3, and Case-4 are given in red, blue, purple and yellow, respectively. The three different solid accretion rates are distinguished by different symbols. We also show the \textit{light} initial disk results by a transparent marker, the \textit{medium} disk result with a 0.5 opacity marker, and the \textit{heavy} disk result with a full opacity marker. The \textit{light}, \textit{medium} and \textit{heavy} results of the same model are connected by a line, as we would expect that in intermediate disk would produce a final H-He fraction approximately along this line. The total masses in the figure are limited to below 15 M$_{\oplus}$, focusing on the distribution for mini-Neptune type planets. Planets which reached the crossover mass (M$_{\textrm{env}}$ = M$_{\textrm{core}}$) are not shown, even if their total mass is below 15 M$_{\oplus}$.
    }
    \label{fig:All_HHe}
\end{figure*}

Figure \ref{fig:All_HHe} summarizes all the nebular H-He mass fractions for the different formation scenarios presented in this work. This fraction is indicated by $f_{\textrm{H-He}}$ which is the H-He mass in the envelope divided by the total mass of the planet. All panels show $f_{\textrm{H-He}}$ as a function of the total mass of the planet. Only masses below 15 M$_{\oplus}$ are shown to focus on mini-Neptune planets and because post-runaway gas accretion was not taken into account in our simulations. Planets which reached the crossover mass or had a total mass above 15 M$_{\oplus}$ were also removed from the figure. Dashed vertical lines reference where $f_{\textrm{H-He}}$ has values of 10\%, 1\% and 0.1\%.\\
The left panels show $f_{\textrm{H-He}}$ when the planet forms at 3 au. All the planetesimal cases (rapid and oligarchic) remain pre-runaway up to 10 Myr with the exception of Case-2 and Case-4 of rapid growth in a \textit{heavy disk}. Rapid growth leads to larger masses and larger values for $f_{\textrm{H-He}}$ than oligarchic growth. This difference in composition between the two formation models is most visible at 3 Myr. If formation times are longer and there is strong envelope enrichment (Case-2) then oligarchic growth can create planets with total masses and H-He mass fractions that overlap those of rapid growth. Overall rapid growth can create planets of masses 2 - 9 M$_{\oplus}$ with H-He envelope fractions of 0.03 to 0.5 at 3 au. Oligarchic growth creates planets with total masses of 0.5 to 4 M$_{\oplus}$ with $f_{\textrm{H-He}}$ values of 5$\times 10^{-4}$ to 0.1.\\
At 5 au there are not as many datapoints for rapid growth as the planets are likely to reach the crossover mass quickly. None of the \textit{heavy disk} cases remain. The planet forming in a \textit{medium disk} under Case-1 remains pre-runaway at 5 Myr and for Case-3 this is past 3 Myr. Planets forming by rapid growth with Case-2 can only do so in a \textit{light disk} in 3 Myr. The oligarchic cases all remain pre-runaway and have smaller H-He mass fractions than at 3 au. This is because the planets forming at 5 au have smaller accretion radii due to a smaller Bondi radius. At some point the accretion radius becomes dominated by the Hill radius, which increases with distance. This will lead to planets at 5 au holding more massive envelopes than those at 3 au for the same total planetary mass. However for the oligarchic growth cases this transition happens too late to see reflected in H-He mass fractions at 3, 5 or 10 Myr.\\
Pebble accretion only forms mini-Neptune planets when there is a \textit{light disk}, which is assumed to coincide with a late formation time (relative to a heavier disk). Furthermore at 3 au a mini-Neptune can only form when Case-3 enrichment applies if formation lasts longer than 5 Myr. In Case-3 the total mass and $f_{\textrm{H-He}}$ stay within the same region as the planetesimal accretion models. Within 3 Myr a 13 M$_{\oplus}$ planet can also form by pebbles with so that $f_{\textrm{H-He}}$=0.1 assuming Case-1 or Case-4. In Case-2 there are no pre-runaway planets even after 3 Myr.\\
At 5 au it is easier to form small planets by pebbles, although still exclusively for the \textit{light disk}. This is contrary to planetesimal formation which favours smaller planets at 3 au. After 3 Myr there is not yet a distinction between any of the enrichment models for pebble formation as they all lead to a planet of 2.5 M$_{\oplus}$ and a $f_{\textrm{H-He}}$ of 0.004. After 10 Myr only Case-1 and Case-3 have remained pre-runaway.

\subsection{Envelope metallicities after formation}
\begin{figure}
    \centering
    \includegraphics[width=\hsize]{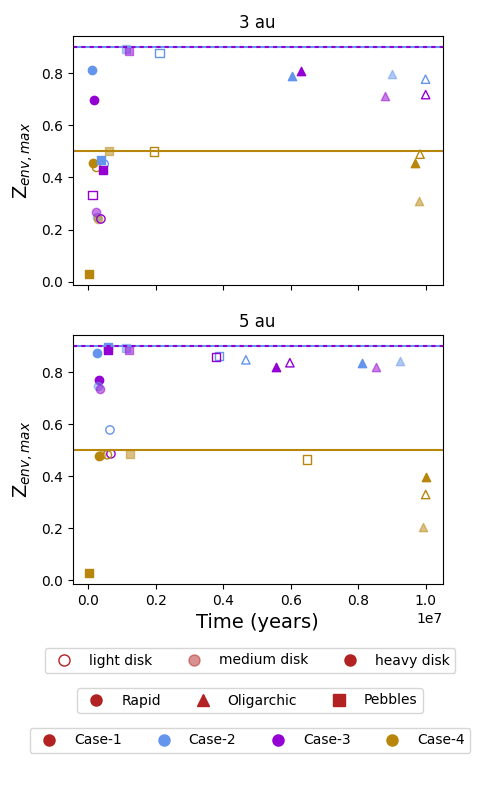}
    \caption{The maximum envelope metallicity that is reached during planetary formation (Z$_{\textrm{env, max}}$) and the time at which this maximum metallicity occurs. The horizontal lines indicate the imposed maximum metallicity for supercritical water (Z$_{\textrm{max}}$ in Table \ref{tab:cases})}
    \label{fig:maxZenv_time}
\end{figure}

\begin{figure}
    \centering
    \includegraphics[width=\hsize]{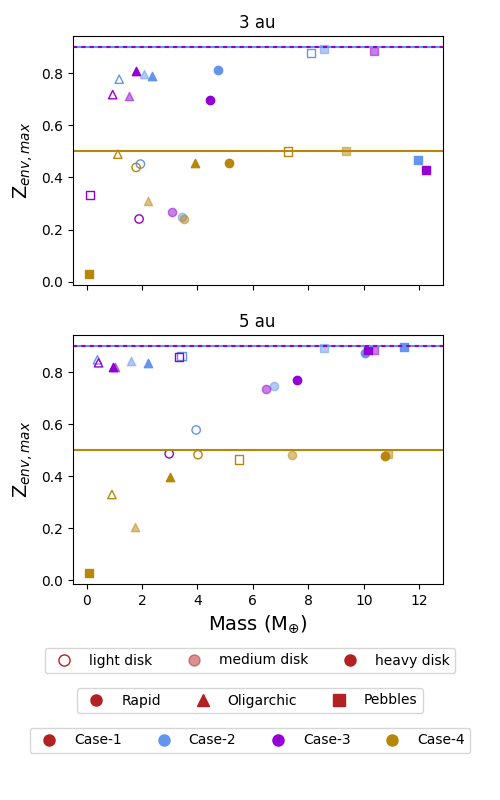}
    \caption{Same data as in Figure \ref{fig:maxZenv_time} but instead of the time, the total mass of the protoplanet is shown when the maximum envelope metallicity is reached.}
    \label{fig:maxZenv_Mcore}
\end{figure}

In Figure \ref{fig:maxZenv_time} and \ref{fig:maxZenv_Mcore} the maximum envelope metallicities (Z$_{\textrm{env, max}}$) are shown for the same set of models as in Figure \ref{fig:All_HHe}, with the exception of Case-1 which by definition evolves to a zero metallicity envelope. Horizontal lines indicate the maximum imposed metallicity for supercritical layers, Z$_{\textrm{max}}$.\\
The rapid growth always has the maximum metallicity occuring  before 400 kyr, which is significantly shorter than the shortest considered formation time of 3 Myr. It is therefore unlikely that rapid growth at 3 or 5 au can create mini-Neptune planets with very high metallicity envelopes, i.e. envelope metallicities that are close to Z$_{\textrm{max}}$.\\
On the other hand, oligarchic growth has maximum metallicities occuring very late, between 4.6 and 10 Myr. These correspond to total masses of 0.5 to 3 M$_{\oplus}$. The oligarchic cases never reach saturation where the envelope metallicity is that of Z$_{\textrm{max}}$.\\
The pebble cases show a wider spread in times when Z$_{\textrm{env, max}}$ is reached. For most of the pebble cases, we find that the maximum metallicity occurs before 3 Myr, but could be delayed to 4 Myr or even 7 Myr if the disk is light.

\section{Discussion} \label{sec:discussion}

\subsection{The effect of mixing and location of deposit}
\begin{figure*}
    \centering
    \includegraphics[width=0.9\hsize]{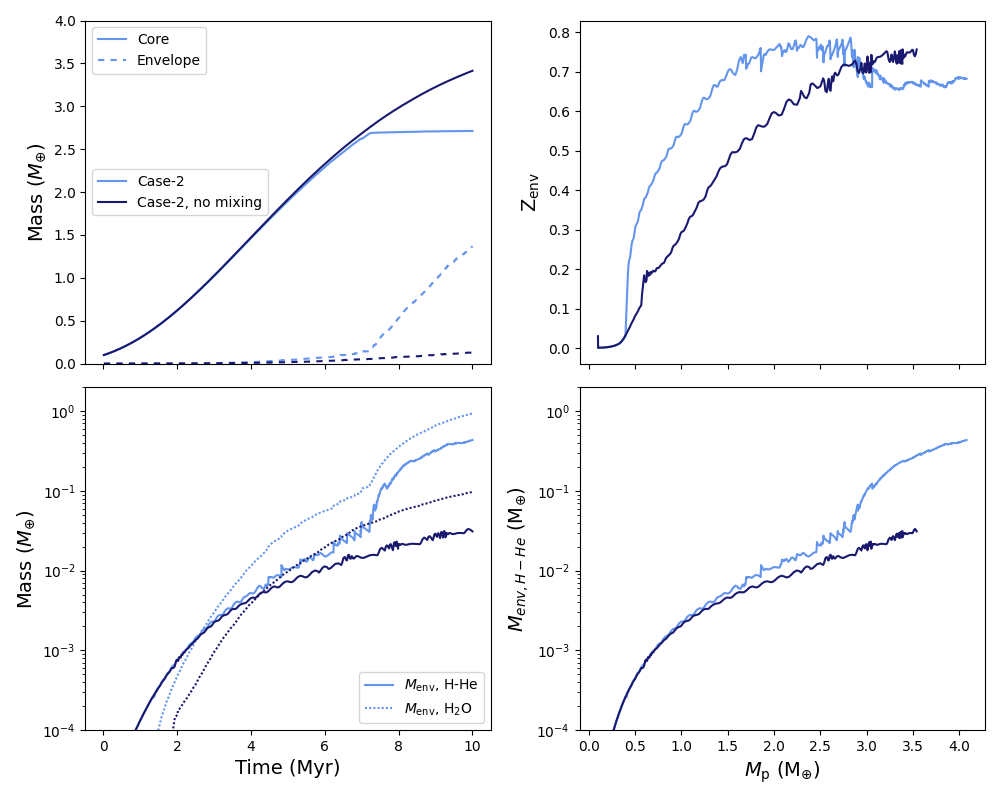}
    \caption{Oligachic growth at 3 au in a \textit{heavy disk}. The light blue line indicates Case-2 without mixing and the dark blue line is for the same conditions but with mixing ($\alpha_{\text{mlt}} = 0.1$). When mixing is included, it  distributes the H-He through the envelope. As our method replaces H-He with water, mixing increases the efficiency of  water deposition. As a result,  in the mixing cases there are cases where all the water can be deposited in the envelope. This reduces the accretion luminosity which leads to acceleration of the envelope's growth.} 
    \label{fig:mixing_1}
\end{figure*}

\begin{figure*}
    \centering
    \includegraphics[width=0.9\hsize]{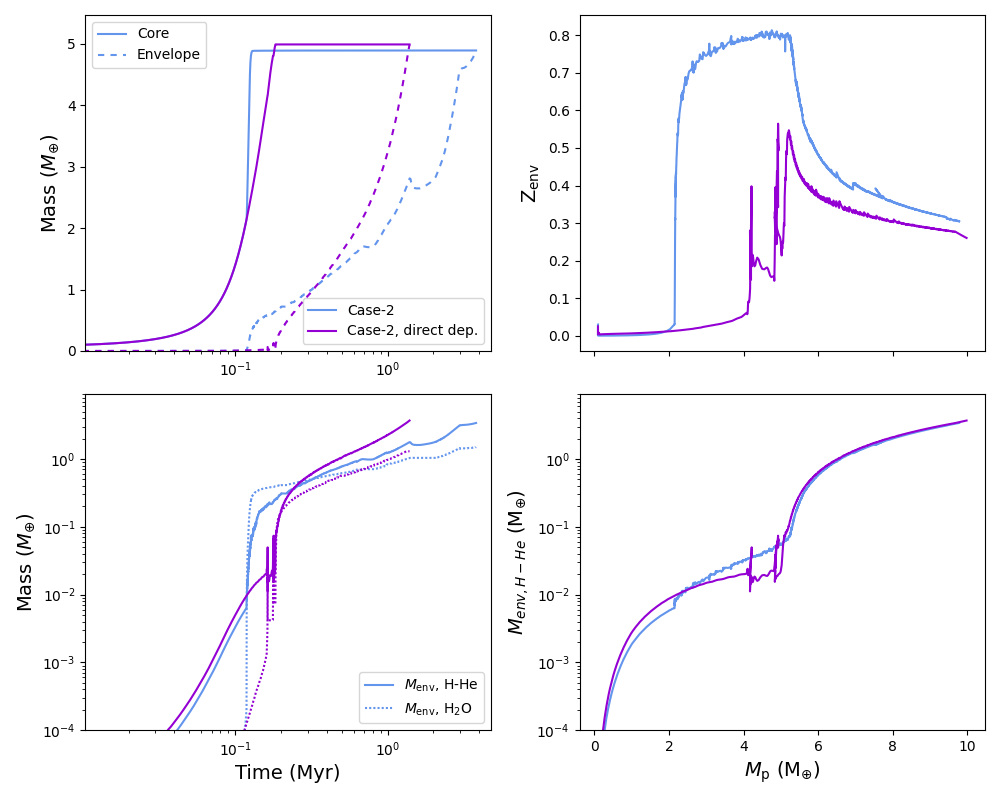}
    \caption{Rapid growth at 3 au in a \textit{heavy disk}. The light blue line shows a homogeneous deposit of heavy elements, which is the default in our model. The purple line shows how the results change when the heavy elements are directly deposited at the relevant radial distance in the envelope.}
    \label{fig:dep_1}
\end{figure*}

The results presented above assumed a homogeneous composition of the envelope due to convective mixing at layers where the Ledoux criterium is met. While it is expected that there is some mixing in protoplanetary envelopes, it is unclear how efficient mixing is. Simulating the planetary formation with MESA allows to include mixing via the mixing length theory (mlt). This, however, requires the knowledge of a dimensionless parameter $\alpha_{\text{mlt}}$ (see Section 2.4 in \citet{Valletta_2020}). For the formation of Jupiter \citet{Valletta_2020} adapted $\alpha_{\text{mlt}} = 0.1$ \citep{Vazan_2015, Muller_2020b} and we use the same mixing in our nominal model. In this Subsection we investigate the effect of mixing by considering a model in which mixing is inhibited.\\
The effect of mixing on the protoplanets core and envelope mass and composition are shown in Figure \ref{fig:mixing_1} for the oligarchic growth at 3 au for Case-2. The initial solid supply is \textit{heavy}. When mixing is included this increases the efficiency of the deposit of heavy elements. The mixing distributes the H-He through the envelope, so that there are more layers where the maximum metallicity is not met. This allows for a larger overall deposit of heavy elements. As a consequence the \textit{mixing} model reaches a point where $f_{\textrm{abl}}$ reaches 1 after 7 Myr. The accretion luminosity becomes zero and gas accretion increases. The final H-He envelop mass is an order of magnitude larger and the core mass is 0.5 M$_{\oplus}$ smaller.\\
\\
We also investigate the effect of our assumption of a homogeneous deposition of heavy elements in the envelope. We  compare the homogeneous deposit to the \textit{direct deposit} as defined in Section \ref{sec:method_enrich}.\\
Figures \ref{fig:dep_1} shows the differences between these deposit models for Rapid growth at 3 au, Case-2. The nominal, \textit{homogeneous deposit} is the same as shown in Figure \ref{fig:growth_rapid}. In the case of \textit{direct deposit}, is takes longer for the envelope to start growing significantly. This is because initially only ablation occurs, which means that in the \textit{direct deposit} case there are only heavy elements in the lower layers. In the \textit{homogeneous deposit} case, water is added to the outer layers as well, so that the density increases and that causes fragmentation to occur more quickly (see Equation \ref{eq:material_strength} and \ref{eq:fragmentation}). Due to this fragmentation almost no solids reach the core so that the accretion luminosity from solid accretion decreases and gas accretion increases. The model with a \textit{direct deposit} of solids reaches fragmentation later. Nevertheless, when both models have reached fragmentation the gas accretion is more efficient in the \textit{direct deposit}. In that case the mass can be deposited high in the envelope and 'trickle down' to the lower layers. The crossover mass is then reached after 1.39 Myr instead of 3.81 Myr.\\
A realistic deposition profile of heavy elements should lie somewhere in between the two extreme cases that we  considered in this work. 
Assessing the physical importance of mixing on planet formation, in combination with envelope enrichment,  would require the following improvements. First, the one-dimensional deposit profile needs to be smoothed appropriately to account for the three-dimensional process \citep{Mordasini_2017}. In the case of planetesimal accretion this initial deposit profile would also need to be improved upon by using a more realistic size distribution \citep{Kaufmann_2023}. Second, the treatment of envelope metallicity should be improved. In this work, the accreted heavy-element mass was added by increasing the metallicity and changing the energy in the relevant layers. However, mass in every layer was conserved during enrichment. Future work should treat mass deposition and envelope enrichment self-consistently by allowing this process to directly change the mass of the relevant layers. \\

\subsection{Primordial envelopes and habitability}
Planets with a primordial, H-He dominated envelope have received increased attention as potentially habitable candidates. The collision-induced absorption of hydrogen can act as a greenhouse effect and thereby create temperate surface conditions \citep{Stevenson_1982, Pierrehumbert_2011, Madhusudhan_2021, Mol_Lous_2022}.
\citet{Madhusudhan_2021} coined the term 'Hycean planets', which host liquid water underneath a hydrogen-dominated atmosphere.
It remains uncertain, however, whether any of the currently observed transiting exoplanets orbit in what can be considered the 'Hycean habitable zone'. The role of a runaway greenhouse effect \citep{Pierrehumbert_2023, Innes_2023} and atmospheric escape \citep{Wordsworth_2012, Mol_Lous_2022} could move the inner habitable zone boundary in comparison to an Earth-like planet.\\
Another open question regarding Hycean planets is whether such planets can accrete the required   amount of H-He to enable these temperate surface conditions in the first place. 
In \citet{Mol_Lous_2022} we showed that planets of sizes 1 - 10 M$_{\oplus}$ that orbit around a sun-like star could host temperate conditions if they are beyond 2 au. Their primordial, pure H-He envelope could be of masses 10$^{-4}$ - 10$^{-5}$ M$_{\oplus}$ at this distance, but more massive when further out. 
The results presented in Figure \ref{fig:All_HHe} show that H-He envelope below 10$^{-3}$ are generally difficult to form, and most of the formed planets consist of much larger H-He mass fractions. The smallest H-He envelopes are formed by oligarchic growth at 5 au. After 3 Myr years these planets have values for M$_{\textrm{env, H-He}}$ ranging from 2.7 $\times 10^{-5}$ M$_{\oplus}$ ($f_{\textrm{H-He}}= 10^{-4}$ when the total mass is 0.27 M$_{\oplus}$) up to 2.24 $\times 10^{-4}$ M$_{\oplus}$ ($f_{\textrm{H-He}}=4 \times 10^{-4}$ when the total mass is 0.56 M$_{\oplus}$). These planets are smaller than those considered in \citet{Mol_Lous_2022}, but still massive enough to hold onto a H-He envelope at 5 au \citep{Mordasini_2020}. \\
We therefore conclude that the formation of H-He envelopes which provide temperature surface conditions is probably rare but possible when the planet forms beyond the ice-line and the formation timescale is long.

\section{Caveats} \label{sec:caveats}
\subsection{Envelope-core interactions and outgassing} \label{cav:outgassing}
Our model does not include interactions between the envelope and the core. We assume that all the nebular hydrogen and helium remain in the envelope and that none is sequestered in the core, to be outgassed at later stages. For (super-)Earth this can play an important role in the development of the atmosphere after the gas disk has disappeared \citep{Elkins-Tanton_2008, Schaefer_2010}. As silicates in the core are expected to be in the magma phase there should be a high solubility of hydrogen \citep{Hirschmann_2012}. That hydrogen would over time be outgassed and replenish the envelope \citep{Chacan_2018}, but that would be accompanied with the atmospheric escape of mostly hydrogen. There could also be a later increase in atmospheric hydrogen if metal-rich impactors oxidate \citep{Genda_2017}. Thus, some of the H-He mass calculated in this work could be stored in the core and released gradually.\\
The envelope-core interactions for H$_{2}$O, not considered in this work, should also be mentioned. While we focus on predicting the mass fraction of H-He after disk accretion, the treatment of water can be improved, which could lead to different results. Water can also be stored efficiently in a magma ocean and outgassed later \citep[e.g.,][]{Dorn_2021, Bower_2022, Sossi_2023}, which can increase the envelope's metallicity after formation.

\subsection{Ablation and fragmentation of silicates}
In this work we only modelled the effect of water enhancement on the envelope. It is clear that rocky material can also ablate and fragment. This is especially the case for pebbles \citep{Brouwers_2018, Brouwers_2020, Steinmeyer_2023}, but also for planetesimals \citep[e.g.]{Bodenheimer_2018}. Similar to  water enrichment, the enrichment with silicates on the one hand increases the mean molecular weight and promotes gas accretion and on the other hand enhances the opacities in the envelope \citep{Ormel_2014, Mordasini_2014, Menou_2023}. The enrichment of silicates alone can create a composition gradient which inhibits convection \citep{Ormel_2021, Markham_2022}. This silicate enrichment has an effect on the long-term evolution of mini-Neptunes and not considering this effect can lead to overpredictions of H-He mass fractions in observed planets \citep{Misener_2022, Vazan_2023}.\\
Future work should consider the enrichment of both water and silicates in the envelopes of protoplanets. However, accounting for both species will introduce more free parameters concerning the mixture of ice and silicates in the solid accretion rate.

\subsection{Limitations of a one-dimensional model} \label{caveats:one_d}
The assumption of spherically symmetric gas and solid accretion that comes with a one-dimensional model does not accurately reflect the reality and thus there are some limitations. First of all there is the recycling of gas that occurs in the outer regions of the accreting envelope. The nebular gas from the disk has a higher entropy than the already accreted gas in the envelope and will mix \citep{Ormel_2015}. This will delay the cooling and contraction, thus prolonging Phase II of gas accretion.
Recycling is an important aspect to planet formation. It can significantly delay the formation timescale which can help to explain the presence of super-Earths and mini-Neptunes, when one-dimensional models would have predicted a transition into runaway gas accretion. This could also possibly help with the formation of Uranus and Neptune \citep{Eriksson_2023}. Three-dimensional gas accretion simulations remain computationally expensive. Recently \citet{Bailey_2023} found a more optimistic comparison between three- and one-dimensional models. They suggest that one-dimensional models can improve their accuracy by reducing the accretion radius to 0.4 times the Bondi radius and considering two distinct outer recycling layers.\\
A second limitation revolves around the deposition profile of heavy elements. In this work we have considered two extremes. In the nominal case we deposited the heavy elements homogeneously and alternatively we solved the deposition profile in the one-dimensional case and deposit the solids accordingly. The latter is realistic if the solid accretion rate is isentropic and the timescales of the impacts is shorter than the azimuthal mixing \citep{Mordasini_2017}.

\subsection{In-situ formation}
The effect of migration is neglected in this work although in-situ formation is rather unrealistic. One origin theory of mini-Neptunes is that they form around the ice-line and migrate inwards once they reach a critical mass \citep{Kuchner_2003, Venturini_2020, Huang_2022, Burn_2024}. The precise value of this critical mass remains uncertain \citep{McNally_2019, Paardekooper_2023} though derivations of it can be found, such as in \citet{Emsenhuber_2023}. 
This formation scenario would naturally lead to a diversity in mini-Neptunes \citep{Bean_2021}.\\
In the rapid growth case some planets in our simulations deplete their feeding zone and enter Phase II gas accretion. We predict that if migration is included this will lead to larger planets.
However, our conclusion that different treatments of solid-envelope interactions can heavily influence the outcome of planet formation is robust.


\section{Summary and Conclusions} \label{sec:concl}
We simulated planet formation assuming different solid accretion rates, and calculated the corresponding gas accretion rate self-consistently. 
The planetary formation locations were set to be outside of the ice-line, where the observed mini-Neptunes could have formed before migrating to shorter orbital distances. 
Our study clearly shows that the assumptions used by planet formation models play a key role in determining the planetary growth history and therefore also the planetary mass and composition. 
Our key conclusions can be summarized as follows:
   \begin{itemize}
        \item The assumptions of the interaction between solids and the planetary envelope strongly affect the planetary growth and can change the primordial gas mass fraction by up to a factor of 10. Nevertheless, we have also identified  cases where this interaction has no or only little influence on the forming planet.  In the case of oligarchic growth, the envelope  remains small and, although it is metal-rich, it does not alter the rest of the formation process. 
        \item Forming mini-Neptunes at 3 au is challenging with pebble accretion due to the high accretion rates. Their formation via pebble accretion becomes more likely at 5 au when the initial disk is relatively light. However, we place more caveats to our pebble result compared to the planetesimals because (1) pebble growth strongly depends on the start of the growth, and (2) pebbles could sublimate before reaching the accretion radius, which might influence the results. 
        \item The impact of envelope pollution is complex. The most extreme solid-envelope interaction cases (Case-1 and Case-2) do not automatically lead to the most extreme outcomes. On the contrary, we find that our Case-4, which considers half of the solids to interact, does not necessarily lead to planets in between the more extreme cases. For example, with Rapid growth at 5 au we find that Case-4 leads to the smallest planets for a given time.
        \item Envelopes of protoplanets during Phase I and II of gas accretion can be dominated by heavy elements.
        \item Our results are consistent with  the observed diversity of  exoplanets \citep[e.g.,][]{Jantof-Hutter_2019}. Variations in the formation location, solid material in the protoplanetary disk, the composition of the solid material, and the formation timescale determine the final mass and composition of the forming planets. We find that $f_{\textrm{abl}}$ can span a range of several order-of-magnitudes and that the envelope's metallicity can range from 0 to full saturation. Our results clearly imply  that small- to intermediate- mass planets should be diverse in terms of mass and composition, depending on the exact formation conditions and growth history.
   \end{itemize}
We find that gas accretion models that include envelope-solids interactions can significantly influence planet formation even for small planetary masses. This has important effects on our understanding of the formation of mini-Neptunes. We note that assuming pebble or planetesimal accretion exclusively could lead to an overestimation of cases which enter runaway gas accretion. \citet{Kessler_2023} showed that giant planet formation can be suppressed when both pebbles and planetesimals are considered.\\
Although the topic is still being investigated, it is often assumed that the observed mini-Neptunes and super-Earths have a large water mass fraction \citep{Venturini_2020, Luque_2022}.
This is further supported by the observation of volatile-rich planets, e.g. in Kepler 138-c and -d \citep{Piaulet_2023}. Furthermore, there are observations indicating the presence of atmospheric water vapor \citep[e.g.,][]{Mikal-Evans_2023}. However, detecting atmospheric water is challenging due to the possible formation of clouds. Water signatures can also overlap with those of methane \citep{Bezard_2022}. For K2-18b, which became notorious as the first mini-Neptune detected with water vapor in the atmosphere \citep{Benneke_2019b}, JWST data have confirmed that the measured signature was due to methane, and not water \citep{Madhusudhan_2023}. For highly-radiated planets JWST should be able to better constrain the volatile abundances \citep{Ancuna_2023, Piette_2023}. If future observations can confirm that mini-Neptunes and super-Earths have water-rich atmosphere, it would support the idea that they have formed beyond the ice-line and migrated inward, as we assumed here.
Other explanations for water-rich atmospheres of small planets could be a late volatile delivery \citep{Elkins-Tanton_2008} or in-situ water formation \citep{Kite_2021}.\\
Our results demonstrate that primordial gas accretion rates are not simple. Assumptions in the solid-envelope interaction, the solid accretion rate and formation location can greatly influence the fraction of H-He after formation. These assumptions as well as aspects not considered in this work (migration, grain opacities) will need to be included in order to explain the observed mini-Neptune and super-Earth population.

\begin{acknowledgements}
      We thank Simon M\"uller for contributing to the re-writing of the original code used in \citet{Valletta_2020} and for helpful discussions. We also thank the referee for their valuable comments. This work has been carried out within the framework of the National Centre of Competence in Research PlanetS supported by the Swiss National Science Foundation under grants 51NF40\_182901 and 51NF40\_205606. The authors acknowledge the financial support of the SNSF.
\end{acknowledgements}

\begin{appendix}

\section{Overview of solid accretion rates} \label{ap:solid_rates}

\subsection{Rapid Growth} \label{app:rapid}
Here we give an overview of how the solid accretion rates were calculated.
In the case of rapid growth we calculate the solid accretion rate, $\dot{M}_{Z}$, from \citet{Pollack_1996}, Eq. 1:
\\
  \begin{equation} \label{eq:Mzdot_rapid}
      \dot{M}_{\text{Z}} =F_{\mathrm{g}} \; \Omega_{\mathrm{k}} \pi R_{\mathrm{cap}}^{2} \Sigma_{\text{Z}} 
  \end{equation}
  \\
which uses the gravitational focusing factor $F_{\mathrm{g}}$. This is calculated following \citet{Greenzweig_1992} (analytical approximation from Appendix B). $\Omega_{\mathrm{k}}$ is the orbital frequency. $\Sigma_{\text{Z}}$ is the solid surface density, described in \ref{subsubsec:ssd}).\\
The capture radius ($R_{\mathrm{cap}}$ or 'enhance radius') is calculated as in \citet{Inaba_2003}, by considering the drag of the already accreted envelope. Using the assumed radius ($r_{\text{pl}}$) and density ($\rho_{\text{pl}}$) of the planetesimals, we find the capture radius $R_{\mathrm{c}}$ in the envelope where it holds that:
\\
\begin{equation} \label{eq:R_cap}
        r_{\mathrm{pl}}=\frac{3}{2} \frac{\rho\left(R_{\mathrm{c}}\right)}{\rho_{\mathrm{pl}}} R_{\mathrm{Hill}},
\end{equation}
\\
which also uses the Hill radius of the protoplanet ($R_{\mathrm{Hill}}$). When there is no sufficient envelope accreted, the capture radius is reduced to the core radius.\\

\subsection{Oligarchic growth} \label{app:olig}
Oligarchic growth is based on the idea that the planetary embryo can already be massive enough to perturb the planetesimals, increasing the temperature and reducing the solid accretion \citep{Ida_1993}. It leads to much longer formation timescales than rapid growth. We adapt the accretion rate from \citet{Fortier_2007}, Eg. 10:
\\
\begin{equation}
    \dot{M}_{Z} = F \frac{\Sigma}{2h} \pi R_{\text{cap}}^{2} v_{\text{rel}}, 
\end{equation}
\\
where $F$ is an efficiency factor that needs to compensate the fact that the accretion rate is underestimated when the planetesimals are assumed to have eccentricities and inclinations all equal to the rms. In \citet{Greenzweig_1992} it is estimated to be $\approx 3$.\\
Following \citet{Fortier_2007} we approximate the inclinations and eccentricities as:
\begin{equation}
    e \approx 2 \dot i
\end{equation}
and find $e$ by Equation 10 in \citet{Thommes_2003}.

\subsection{Solid surface density}\label{subsubsec:ssd}
The initial solid surface density, $\Sigma_{\text{Z, 0}}$, is calculated by Equation \ref{eq:Sigma_init}.
The planet accretes solid materials from its feeding zone. The extend of this zone reaches $\sqrt{12 + (e a_{\mathrm{orb}} / R_{\mathrm{Hill}})^{2}} R_{\text{Hill}}$ in both the inner and outer direction, following \citet{Pollack_1996}. Therefore the inner and outer radii of the feeding zone are given by:
\\
\begin{equation}
    R_{\mathrm{in}} = a - \sqrt{12 + (e a_{\mathrm{orb}} / R_{\mathrm{Hill}})^{2}} \; R_{\text{Hill}} , 
\end{equation}
\begin{equation}
    R_{\mathrm{out}} = a + \sqrt{12 + (e a_{\mathrm{orb}} / R_{\mathrm{Hill}})^{2}} \; R_{\text{Hill}}, 
\end{equation}
\\
where $e$ is the planetesimal's  eccentricity. In the case of rapid growth this is calculated by Equation 6 in \citet{Pollack_1996}.\\
While the solid surface density is expected to be lower at $R_{\mathrm{in}}$ and higher at $R_{\mathrm{out}}$ compared to $\Sigma_{\text{Z, 0}}$, this will partly cancel out such that the difference is negligible.\\
$\Sigma_{\text{Z}}$ does reduce when solid material is accreted onto the planet. We calculate $\Sigma_{\text{Z}}$ by:
\\
\begin{equation}
    \Sigma_{\text{Z}} = \Sigma_{\text{Z, 0}} - \frac{M_{\text{acc}}}{\pi (R_{\mathrm{out}}^{2} - R_{\mathrm{in}}^{2})},
\end{equation}
\\
with $M_{\text{acc}}$ being  the already accreted solid material.\\
Not considered are the scattering of planetesimals by the growing protoplanet \citep{Zhou_2007, Shiraishi_2008} which would decrease the amount of solid material.

\subsection{Pebble Accretion} \label{app:pebbles}
The solid accretion rate of pebbles is adapted from \citet{Lambrechts_2014}, Eq. 31. It is given by:
\\
    \begin{equation} \label{eq:Mzdot_peb}
        \begin{aligned}
\dot{M}_{\mathrm{Z}} \approx & 4.8 \times 10^{-6}\left(\frac{M_{\text{c}}}{\mathrm{M}_{\oplus}}\right)^{2 / 3}\left(\frac{Z_{0}}{0.01}\right)^{25 / 18}\left(\frac{M_{*}}{\mathrm{M}_{\odot}}\right)^{-11 / 36}\left(\frac{\beta}{500 \mathrm{~g} \mathrm{~cm}^{-2}}\right) \\
& \times\left(\frac{r}{10 \mathrm{ au}}\right)^{-5 / 12}\left(\frac{t}{10^{6} \mathrm{yr}}\right)^{-5 / 18} \left( \frac{\dot{M}_{\mathrm{Z, 3D}}}{\dot{M}_{\mathrm{Z, 2D}}}\right ) \mathrm{M}_{\mathrm{E}} \mathrm{yr}^{-1}, 
\end{aligned}
\end{equation}
\\
where $Z_{0}$ is the disk's  metallicity and it is set to 0.010270. $\beta$ is given in Equation \ref{eq:beta} and $\tau_{\mathrm{disk}} = 3 \times 10^{6} \mathrm{~yrs}$.\\
Furthermore there is a multiplication with a 3D factor from \citet{Venturini_2017} which slightly reduces the mass accretion rate. This 3D factor is:
\\
\begin{equation}
    \frac{\dot{M}_{\mathrm{Z, 3D}}}{\dot{M}_{\mathrm{Z, 2D}}} = \left(\frac{\pi\left(\tau_{f} / 0.1\right)^{1 / 3} R_{\text{Hill}}}{2 \sqrt{2 \pi} H_{\mathrm{peb}}}\right), 
\end{equation}
\\
 with $H_{\mathrm{peb}}=H_{\mathrm{gas}} \sqrt{\alpha / \tau_{\mathrm{f}}}$. $\alpha$ is the disk's  viscosity which is set to  $\alpha = 10^{-4}$.
$H_{\mathrm{gas}}$ is the scale-height of the gas disk $H_{\mathrm{gas}}=c_{\text{s}} / \Omega$. This is approximated as $(h=H / r)$ with $h(r)=h_{0}(r / A U)^{1 / 4}$. In standard case $h_{0}=0.036$. See \citet{Venturini_2017} Section 2.2 for this. Finally, in \citet{Lambrechts_2014} Equation 20:
\\
\begin{equation}
    \tau_{\mathrm{f}} \approx \frac{\sqrt{3}}{8} \frac{\epsilon_{\mathrm{p}}}{\eta} \frac{\Sigma_{\mathrm{p}}}{\Sigma_{\mathrm{g}}},
\end{equation}
\\
where we use $\epsilon_{\text{p}}$=0.5 and $\eta = 0.0015 \times \left ( a / 1 \text{au} \right )$
Pebble accretion stops when the pebble isolation mass (M$_{\textrm{iso}}$) is reached. Based on \citep{Lambrechts_2014b} we use that:
\\
\begin{equation}
    M_{\textrm{iso}}  = 20 \left ( \frac{a}{5 \textrm{ au}} \right )^{\frac{3}{4}} \, M_{\oplus}.  
\end{equation}
\\
It is believed that the pebble accretion abruptly halts once the isolation mass is reached. In this work, however,  we smooth the transition by applying  a reduction factor to the pebble accretion rate. This reduction factor is 1 until the total mass is above half the isolation mass, and linearly reduces in mass, so that it is 0 when the total mass equals the pebble isolation mass.

\section{Fraction of ablation} \label{app:abl_frac}
The fraction of ablating or fragmenting material ($f_{\textrm{abl}}$) is calculated every timestep and changes as the size and composition of the envelope change. We follow the semi-analytical approached derived in \citet{Valletta_2019}, which was based on the mass loss and motion of a planetary envelope derived in \citet{Podolak_1988}. The  interacting solid material is assumed to be water. Therefore, the density of the planetesimals or pebbles is set to be $\rho_{\textrm{pl}}=1 \, \textrm{g} \, \textrm{cm}^{-3}$. 
This is similar to previous assumptions \citep{Venturini_2016, Valletta_2020} that are justified by observation of comets which indicate that water is the mayor volatile specie in the solar-system. Of course in reality, the accreted material is unlikely to be made of pure-water. Below we refer to the impactors as planetesimals although the same method is applied for pebbles.\\
We integrate the planetesimals' position, mass, and velocity from the outer boundary of the envelope model to the inner boundary, or until all the planetesimal mass is ablated or fragmented. The assumed initial velocity and location are $v_{\textrm{init}}=$10 km sec$^{-1}$ and $r_{\textrm{init}}= 2 R_{\textrm{pl}}$, respectively, with $R_{\textrm{pl}}$ being the outer boundary of our model. The planetesimal's velocity at location $r$ inside the envelope is given by:
\\
\begin{equation}
    v_{\textrm{pl}}^{2}=v_{\textrm{init}}^{2}+2 G M_{\text{p}}\left(\frac{1}{r}-\frac{1}{r_{\textrm{init}}}\right).
\end{equation}
\\
We neglect the envelope's mass beyond the location $r$ since the mass is negligible  compared to the core mass and lower part of the envelope. The mass loss at location $r$ is calculated using Equation 14 in \citet{Valletta_2019}:
\\
\begin{equation}
    \frac{\text{d} m_{\text{pl}}}{\text{d} r}=-\frac{A}{Q}\left(C_{\text{h}} \frac{\rho(r) v_{\textrm{pl}}^{2}}{2}+\epsilon \sigma T(r)^{4} \frac{1}{v_{\textrm{pl}}}\right),
\end{equation}
\\
where $m_{\text{pl}}$ is the mass of the planetesimal, $A$ is the area of the planetesimal which naturally decreases as the planetesimal loses mass and $\rho(r)$ is the density at $r$. 
$C_{\text{h}}$ and $\epsilon$ are  efficiency factors for which the appropriate values are uncertain. $C_{\text{h}}$ is the fraction of kinetic energy transferred to the planetesimal. $\epsilon$ is the product of the emissivity of the gas and the planetesimal's impact coefficient. We set both to 0.01. In \citet{Valletta_2019}, $C_{\text{h}}$ and $\epsilon$ were left as free parameters and while their value affects the planetary growth, its effect is smaller in comparison to  other assumptions considered in this work, such as the solid accretion rate and envelope mixing. Unlike in  \citet{Valletta_2019}, here we simplify the calculation of the planetesimal's  trajectory by assuming that it moves straight to the core. In other words, we assume  an impact parameter of 0 and no angular contributions to the velocity. $Q$ is the latent heat caused upon vaporization and is given by:
\\
\begin{equation}
    Q = C_{\text{p}} \times T_{\text{f}} + E_{0}.
\end{equation}
\\
Here $C_{p}$ is the specific heat of water in the liquid phase, set to 4.2$\times 10^7$ erg g$^{-1}$ K$^{-1}$. $E_{0} = 2.8 \times 10^{10}$ erg g$^{-1}$ is the latent heat of vaporisation in the solid phase. The  values are taken from \citet{Pollack_1986} (Table 1). $T_{\text{f}}$ is the difference between the initial temperature and the present temperature, which is 373 Kelvin. \\
The planetesimal can be completely destroyed when two conditions are met \citep{Pollack_1986}. First, if the pressure gradient in the envelope surrounding the planetesimal is larger than the material strength:
\\
\begin{equation} \label{eq:material_strength}
    \frac{1}{2} \rho(r) v_{\textrm{pl}}^{2} \geq S
\end{equation}
\\
with $S$ being the strength of the compressive material, set to 1$\times 10^{6}$ Ba (0.1 Mpa) for ice \citep{Pollack_1979}. Second, if the planetesimal is sufficiently small so that its self-gravity can not prevent  fragmentation. 
\\
\begin{equation} \label{eq:fragmentation}
    r_{\textrm{pl}} < \sqrt{\frac{5 v_{\textrm{pl}}^{2} \rho(r)}{8 \pi G \rho_{\textrm{pl}}^{2}}}
\end{equation}
\\
If both these criteria are met, fragmentation occurs and $f_{\textrm{abl}}$ is set to 1.

\section{Importance of the core's mass-radius (M-R) relation}  \label{subsec:MR}
\begin{figure}
    \centering
    \includegraphics[width=\hsize]{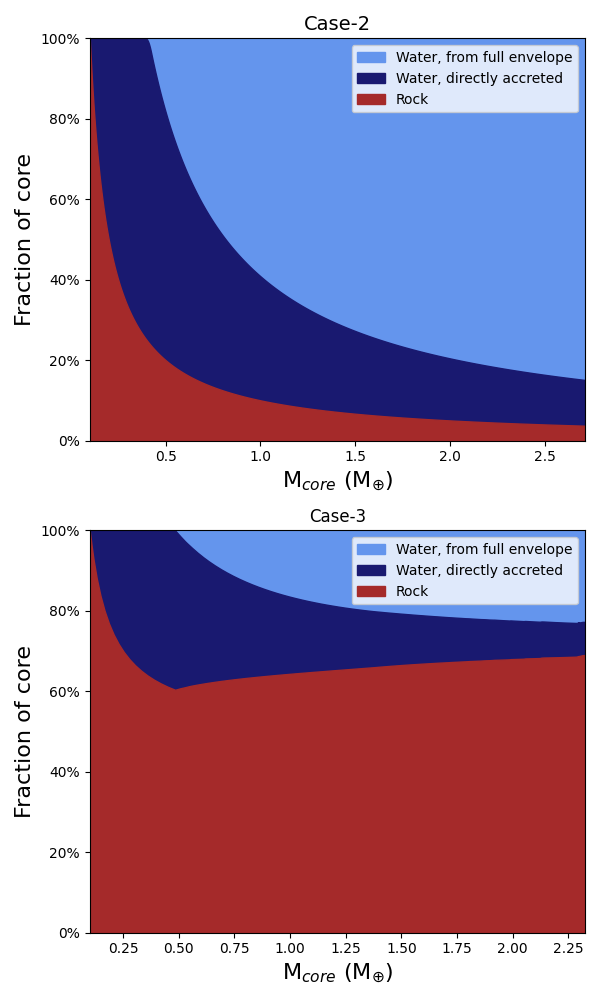}
    \caption{Core composition during oligarchic growth at 3 au with a \textit{heavy} initial disk. The upper panel shows Case-2, where only water is accreted. As a result, the rock mass fraction, which is the initial model, decreases. The lower panel shows Case-3, where 50\% water and 50\% rock is accreted. As all the rock directly reaches the core, but the water can evaporate in the envelop, the core always has a rock mass fraction above 50\%. The water which is added the core can either directly reach the core (dark blue regions) or  be added because the envelope was saturated (light blue).}
    \label{fig:core}
\end{figure}

\begin{figure}
    \centering
    \includegraphics[width=\hsize]{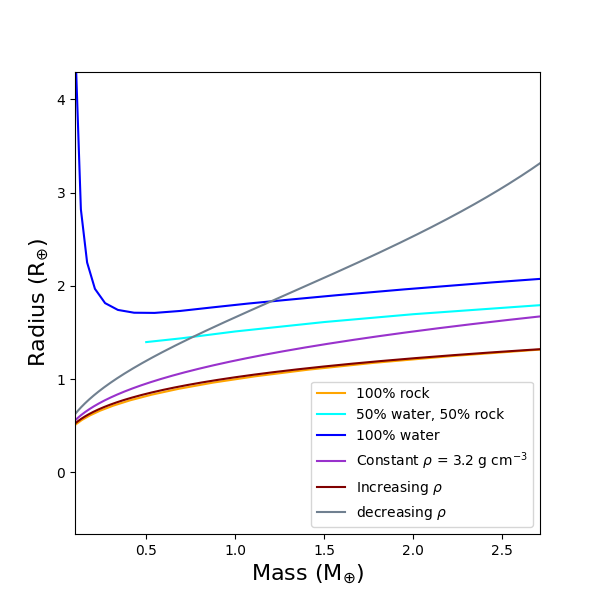}
    \caption{The orange, light blue, and dark blue lines are mass-radius relationships taken from \citep{Zeng_2019}, which are (a) 100\% made of rock of Earth-like composition (i.e. 32.5\% iron and 67.5\% MgSiO3) (b) 50\% made of an Earth-like rocky core and 50\ or (c) 100\% pure H$_{2}$O. The purple line shows the M-R relationship for a constant density of 3.2 g cm$^{-3}$, the  default used  in this work. The \textit{Increasing }$\rho$ model is a fit to the rocky core composition. The \textit{Decreasing }$\rho$ is the minimum decrease in core density which affects the gas accretion rate. Since the gas accretion rate is only altered when the core's density is significantly lower than interior models predict, we conclude that our assumption of a constant core density does not significantly influence the results.}
    \label{fig:MR_rel}
\end{figure}

In our current model the changes in the core's composition are not considered. All the material that reaches the core, whether that is ice or solids, adds to the core mass in the same way. Using the total accreted mass and assuming a constant core density of 3.2 g cm$^{-3}$ we calculate the core radius and this core radius sets the lower boundary in our atmosphere model. A more realistic model would need to infer the core's  radius with an interior model based on the assumed accreted material. We evaluate the influence of the core density on our results by using a gradually decreasing or increasing core density.\\
We do not consider the interaction of rocky material, and for simplicity the rock fraction in the core follows directly from the solid accretion rate. The water/ice mass fraction can have two different sources. First, it was directly accreted, which happens when there is no fragmentation. Second, there could be  water excess since not all the water could be deposited in the  envelope. We also consider this to be part of the core although  in reality this should be added to the envelope mass at the lower layers, creating an ocean. \\
The upper panel in Figure \ref{fig:core} shows the core's composition for oligarchic growth at 3 au with Case-2. From the assumption that in Case-2 there is no rock in the solid accretion rate, the rock fraction of the core only decreases. The lower panel shows the core composition for Case-3 which by definition always has at least 50\% accreted rock directly onto the core. The dark blue represents ice that is directly accreted to the core. The light blue is also water added to the core, but water that did not fit in the envelope and was thus moved to lower layers. For both of these water contributions to the core it is unknown what their thermodynamic properties would be. While the directly accreted water would reach the core in the solid phase, a subsequent impact might still vaporize, therefore also adding water vapor to the envelope. The water which did not fit in the envelope was either because outer layers were too cold for water vapor (small envelope) and thus condensed. It is more common that all layers have reached the maximum metallicity for supercritical water.
Figure \ref{fig:MR_rel} shows the M-R relation of the core with a constant density of 3.2 g cm$^{-3}$. For comparison, we also show the mass-radius relationships of planets taken from \citet{Zeng_2019}. These M-R relations are for planets including their atmospheres and should thus not be directly compared to the M-R of the core. Regardless we apply another core density subscription based on a pure rocky planet, as our aim is merely to determine whether the core's radius can affect the results of our formation model. We find that scaling the core density as $\rho_{\text{c}}$ = 3.4 + $ 2 \sqrt{\text{M}_{\text{c}} / M_{\oplus}}$ gives a similar core radius to the 100\% rock planet ( see \textit{Increasing} $\rho$ in Figure \ref{fig:MR_rel}).\\
Changing the core's density has two competing effects. On one hand, a higher core density leads to a smaller core radius and this core radius is used as a lower boundary in the atmosphere model. Meanwhile the accretion radius stays the same so that the volume which the envelope occupies is slightly larger, thus the envelope can be more massive. This effect, however, is very small since the core radius is about two orders of magnitude smaller than the accretion radius. On the other hand, a higher core density increases the accretion luminosity, which decreases the amount of gas which can fit within the accretion radius.\\
We apply this increasing core density to the formation with oligarchic growth at 3 au under Case-1, as this is the case that assumes the solid accretion is of pure-rock. The final envelope is indeed negligibly larger when using the higher density core, namely 0.082 M$_{\oplus}$ rather than 0.079 M$_{\oplus}$.\\
We do a similar study for the water-rich core in Case-2 but using an arbitrary scaling to the core's density. Rather than trying to simulate the core's M-R realistically, we simply  decrease the density until we see a significant effect in our results. This is achieved when we decrease the density as $\rho_{\text{c}}$ = 3.4 - $ 2 \left ( {M_{c} / \text{M}_{\oplus}} \right )^{1/3}$. As shown in the Figure \ref{fig:MR_rel} this leads to core radii that are more than 1 \Rearth larger than pure water planets. Applying such a low core density to the oligarchic formation at 3 au with Case-2 leads to a core mass of 3.45 M$_{\oplus}$ rather than 2.71 M$_{\oplus}$ and an envelope of 0.3 M$_{\oplus}$ instead of 1.36 M$_{\oplus}$. The primordial envelope mass, $M_{\text{env, H-He}}$ also decreases from 0.43 M$_{\oplus}$ to 0.07 M$_{\oplus}$. This occurs because a low density core eventually leads to a slightly smaller envelope and the fraction of ablation always remains low because there is not enough envelope to replace as water. However, the nominal core density model with a slightly larger envelope does reach a point where most of the solids can enrich the envelope and this causes a great reduction in the accretion luminosity, promoting gas accretion for the final 3 Myrs.\\
The core's mean density that we have applied to a water-dominated core is notably lower than interior models predict (see Figure \ref{fig:MR_rel} and e.g. \citep{Haldemann_2020}). We therefore conclude that an extremely low core density can lead to significantly different primordial envelopes. However, we note that our variations in the core's density is merely a parameter study and is not based on realistic interior models. 
A more realistic core model could be implemented in future work, where it would also need to be considered that rock and water would be mixed in this core \citep[e.g.,][]{Vazan_2022}. 
At the same time, it should be noted that the core's M-R relationship is likely to be less important in comparison to the effect of the chemical interactions between the core and envelope (see Subsection \ref{cav:outgassing}). This would play an important role in determining how the primordial gas is distributed within the planet. 

\section{Result tables} \label{app:results}
Tables \ref{tab:results_rapid}, \ref{tab:results_olig}, \ref{tab:results_pebb} give the composition of the (proto)planet when assuming rapid, oligarchic and pebble growth respectively.
The core mass (M$_{\textrm{core}}$), core water mass fraction (x$_{\textrm{water}}$), envelope mass (M$_{\textrm{env}}$) and envelope metallicity (Z$_{\textrm{env}}$) are given at 3 Myr and 10 Myr. If the crossover mass is reached before 3 Myr, then the planetary composition at the crossover mass is given, rather than at 3 Myr, and no data is given for 10 Myr. If the crossover mass is reached between 3 and 10 Myr, the composition at the crossover mass is given rather than the composition at 10 Myr. The value of x$_{\textrm{water}}$ strongly depends on the assumed solid accretion type, as discussed in Appendix \ref{subsec:MR}.

\begin{table*}[] 
\begin{tabular}{|lll|llll|llll|}
\hline
\multicolumn{3}{|l|}{Rapid growth} & \multicolumn{4}{l|}{After 3 Myr or at M$_{\textrm{core}}$ = M$_{\textrm{env}}$} & \multicolumn{4}{l|}{After 10 Myr or at M$_{\textrm{core}}$ = M$_{\textrm{env}}$} \\ \hline
\multicolumn{2}{|l|}{3 au} & M$_{\textrm{core}}$ = M$_{\textrm{env}}$? & M$_{\textrm{core}}$ (M$_{\oplus}$) & x$_{\textrm{water}}$ & M$_{\textrm{env}}$ (M$_{\oplus}$) & Z$_{\textrm{env}}$ & M$_{\textrm{core}}$ (M$_{\oplus}$) & x$_{\textrm{water}}$ & M$_{\textrm{env}}$ (M$_{\oplus}$) & Z$_{\textrm{env}}$ \\ \hline
\multirow{3}{*}{Heavy} & \multicolumn{1}{l|}{Case-1} & No & 5.64 & 0 & 1.04 & 0 & 6.2 & 0 & 2.74 & 0 \\
 & \multicolumn{1}{l|}{Case-2} & 3.8 Myr & 4.89 & 0.98 & 4.57 & 0.31 & 4.89 & 0.98 & 4.89 & 0.31 \\
 & \multicolumn{1}{l|}{Case-3} & No & 4.58 & 0.41 & 0.98 & 0.14 & 4.83 & 0.39 & 2.49 & 0.1 \\
 & \multicolumn{1}{l|}{Case-4} & 2.5 Myr & 5.15 & 0.98 & 5.15  & 0.22 & n.a. & n.a. & n.a. & n.a. \\ \hline
\multirow{3}{*}{Medium} & \multicolumn{1}{l|}{Case-1} & No & 3.59 & 0 & 0.27 & 0 & 3.73 & 0 & 0.72 & 0 \\
 & \multicolumn{1}{l|}{Case-2} & No & 3.45 & 0.97 & 0.71 & 0.09 & 3.45 & 0.97 & 2.24 & 0.06 \\
 & \multicolumn{1}{l|}{Case-3} & No & 3.38 & 0.46 & 0.43 & 0.11 & 3.45 & 0.45 & 0.82 & 0.09 \\ 
 & \multicolumn{1}{l|}{Case-4} & No & 3.43 & 0.97 & 0.81  & 0.09 & 3.43 & 0.97 & 2.77 & 0.09 \\ \hline
\multirow{3}{*}{Light} & \multicolumn{1}{l|}{Case-1} & No & 1.97 & 0 & 0.09 & 0 & 2.0 & 0 & 0.16 & 0 \\
 & \multicolumn{1}{l|}{Case-2} & No & 1.96 & 0.95 & 0.22 & 0.29 & 1.97 & 0.95 & 0.54 & 0.27 \\
 & \multicolumn{1}{l|}{Case-3} & No & 1.93 & 0.47 & 0.1 & 0.11 & 1.95 & 0.46 & 0.17 & 0.1 \\ 
 & \multicolumn{1}{l|}{Case-4} & No & 1.92 & 0.95 & 0.23  & 0.29 & 1.92 & 0.95 & 0.49 & 0.31 \\ \hline
\multicolumn{2}{|l|}{5 au} &  & M$_{\textrm{core}}$ (M$_{\oplus}$) & x$_{\textrm{water}}$ & M$_{\textrm{env}}$ (M$_{\oplus}$) & Z$_{\textrm{env}}$ & M$_{\textrm{core}}$ (M$_{\oplus}$) & x$_{\textrm{water}}$ & M$_{\textrm{env}}$ (M$_{\oplus}$) & Z$_{\textrm{env}}$ \\ \hline
\multirow{3}{*}{Heavy} & \multicolumn{1}{l|}{Case-1} & 1.98 Myr & 15.5 & 0 & 15.5 & 0 & n.a. & n.a. & n.a. & n.a. \\
 & \multicolumn{1}{l|}{Case-2} & 0.434 Myr & 10.8 & 0.99 & 10.8 & 0.26 & n.a. & n.a. & n.a. & n.a. \\
 & \multicolumn{1}{l|}{Case-3} & 2.83 Myr & 9.38 & 0.3 & 9.38 & 0.1 & n.a. & n.a. & n.a. & n.a. \\ 
 & \multicolumn{1}{l|}{Case-4} & 0.50 Myr & 10.9 & 0.99 & 10.9  & 0.27 & n.a. & n.a. & n.a. & n.a. \\ \hline
\multirow{3}{*}{Medium} & \multicolumn{1}{l|}{Case-1} & 7.06 Myr & 8.18 & 0 & 2.48 & 0 & 10.1 & 0 & 10.1 & 0 \\
 & \multicolumn{1}{l|}{Case-2} & 1.15 Myr & 6.82 & 0.99 & 6.82 & 0.28 & n.a. & n.a. & n.a. & n.a. \\
 & \multicolumn{1}{l|}{Case-3} & 4.73 Myr & 7.13 & 0.42 & 3.63 & 0.11 & 7.63 & 0.39 & 7.63 & 0.08 \\ 
 & \multicolumn{1}{l|}{Case-4} & 1.27 Myr & 7.27 & 0.99 & 7.27  & 0.24 & n.a. & n.a. & n.a. & n.a. \\ \hline
\multirow{3}{*}{Light} & \multicolumn{1}{l|}{Case-1} & No & 4.18 & 0 & 0.44 & 0 & 4.41 & 0 & 1.03 & 0 \\
 & \multicolumn{1}{l|}{Case-2} & 3.43 Myr & 3.95 & 0.97 & 3.06 & 0.27 & 3.95 & 0.97 & 3.95 & 0.25 \\
 & \multicolumn{1}{l|}{Case-3} & No & 3.15 & 0.36 & 0.59 & 0.14 & 3.33 & 0.35 & 1.33 & 0.12 \\ 
 & \multicolumn{1}{l|}{Case-4} & 5.92 Myr & 3.99 & 0.97 & 1.34  & 0.28 & 4.0 & 0.97 & 4.0 & 0.25 \\ \hline
\end{tabular}  \caption{Properties of planets grown by rapid growth at 3 or 5 au. \textit{Heavy}, \textit{Medium} and \textit{Light} refer to the disk models presented in Table \ref{tab:disk}.} \label{tab:results_rapid}
\end{table*}

\begin{table*}[]
\begin{tabular}{|lll|llll|llll|}
\hline
\multicolumn{3}{|l|}{Oligarchic growth} & \multicolumn{4}{l|}{After 3 Myr or at M$_{\textrm{core}}$ = M$_{\textrm{env}}$} & \multicolumn{4}{l|}{After 10 Myr or at M$_{\textrm{core}}$ = M$_{\textrm{env}}$} \\ \hline
\multicolumn{2}{|l|}{3 au} & M$_{\textrm{core}}$ = M$_{\textrm{env}}$? & M$_{\textrm{core}}$ (M$_{\oplus}$) & x$_{\textrm{water}}$ & M$_{\textrm{env}}$ (M$_{\oplus}$) & Z$_{\textrm{env}}$ & M$_{\textrm{core}}$ (M$_{\oplus}$) & x$_{\textrm{water}}$ & M$_{\textrm{env}}$ (M$_{\oplus}$) & Z$_{\textrm{env}}$ \\ \hline
\multirow{3}{*}{Heavy} & \multicolumn{1}{l|}{Case-1} & No & 1.02 & 0 & 2.4 $\times 10^{-3}$ & 0 & 3.53 & 0 & 7.9 $\times 10^{-2}$ & 0 \\
 & \multicolumn{1}{l|}{Case-2} & No & 1.03 & 0.9 & 5.35 $\times 10^{-3}$ & 0.56 & 2.71 & 0.96 & 1.37 & 0.68 \\
 & \multicolumn{1}{l|}{Case-3} & No & 0.86 & 0.36 & 8.5 $\times 10^{-3}$ & 0.7 & 2.32 & 0.31 & 0.16 & 0.76 \\
 & \multicolumn{1}{l|}{Case-4} & No & 1.02 & 0.9 & 5.7 $\times 10^{-3}$  & 0.38 & 3.7 & 0.97 & 0.32 & 0.45 \\ \hline
\multirow{3}{*}{Medium} & \multicolumn{1}{l|}{Case-1} & No & 0.67 & 0 & 9.2 $\times 10^{-4}$ & 0 & 2.20 & 0 & 2.2 $\times 10^{-2}$ & 0 \\
 & \multicolumn{1}{l|}{Case-2} & No & 0.67 & 0.85 & 1.8 $\times 10^{-3}$ & 0.47 & 2.13 & 0.95 & 9.1 $\times 10^{-2}$ & 0.78 \\
 & \multicolumn{1}{l|}{Case-3} & No & 0.64 & 0.40 & 1.6 $\times 10^{-3}$ & 0.48 & 1.61 & 0.35 & 3.5 $\times 10^{-2}$ & 0.67 \\ 
 & \multicolumn{1}{l|}{Case-4} & No & 0.71 & 0.86 & 1.5 $\times 10^{-3}$ & 0.25 & 2.2 & 0.95 & 3.3 $\times 10^{-2}$ & 0.31 \\ \hline
\multirow{3}{*}{Light} & \multicolumn{1}{l|}{Case-1} & No & 0.41 & 0 & 2.9 $\times 10^{-4}$ & 0 & 1.17 & 0 & 4.6 $\times 10^{-3}$ & 0 \\
 & \multicolumn{1}{l|}{Case-2} & No & 0.41 & 0.75 & 1.1 $\times 10^{-3}$ & 0.69 & 1.15 & 0.91 & 2.5 $\times 10^{-2}$ & 0.78 \\
 & \multicolumn{1}{l|}{Case-3} & No & 0.41 & 0.38 & 2.9 $\times 10^{-4}$ & 0.01 & 0.93 & 0.36 & 1.1 $\times 10^{-2}$ & 0.71 \\
 & \multicolumn{1}{l|}{Case-4} & No & 0.41 & 0.76 & 7.3 $\times 10^{-4}$  & 0.4 & 1.12 & 0.91 & 1.6 $\times 10^{-2}$ & 0.47 \\ \hline
\multicolumn{2}{|l|}{5 au} &  & M$_{\textrm{core}}$ (M$_{\oplus}$) & x$_{\textrm{water}}$ & M$_{\textrm{env}}$ (M$_{\oplus}$) & Z$_{\textrm{env}}$ & M$_{\textrm{core}}$ (M$_{\oplus}$) & x$_{\textrm{water}}$ & M$_{\textrm{env}}$ (M$_{\oplus}$) & Z$_{\textrm{env}}$ \\ \hline
\multirow{3}{*}{Heavy} & \multicolumn{1}{l|}{Case-1} & No & 0.55 & 0 & 1.8 $\times 10^{-4}$ & 0 & 2.98 & 0 & 2.3 $\times 10^{-2}$ & 0 \\
 & \multicolumn{1}{l|}{Case-2} & No & 0.55 & 0.82 & 1.0 $\times 10^{-3}$ & 0.77 & 2.89 & 0.97 & 0.13 & 0.82 \\
 & \multicolumn{1}{l|}{Case-3} & No & 0.49 & 0.34 & 1.0 $\times 10^{-3}$ & 0.81 & 1.95 & 0.33 & 0.04 & 0.79 \\ 
 & \multicolumn{1}{l|}{Case-4} & No & 0.56 & 0.82 & 2.2 $\times 10^{-4}$  & 0.14 & 2.97  & 0.97 & 4.5 $\times 10^{-2}$ & 0.39 \\ \hline
\multirow{3}{*}{Medium} & \multicolumn{1}{l|}{Case-1} & No & 0.39 & 0 & 7 $\times 10^{-5}$ & 0 & 1.77 & 0 & 5.8 $\times 10^{-3}$ & 0 \\
 & \multicolumn{1}{l|}{Case-2} & No & 0.39 & 0.74 & 2.1 $\times 10^{-4}$ & 0.66 & 1.73 & 0.94 & 4.4 $\times 10^{-2}$ & 0.83 \\
 & \multicolumn{1}{l|}{Case-3} & No & 0.37 & 0.33 & 3.7 $\times 10^{-4}$ & 0.78 & 1.23 & 0.32 & 1.4 $\times 10^{-2}$ & 0.81 \\ 
 & \multicolumn{1}{l|}{Case-4} & No & 0.38 & 0.74 & 7 $\times 10^{-5}$  & 0.06 & 1.77 & 0.94 & 6.4 $\times 10^{3}$ & 0.21 \\ \hline
\multirow{3}{*}{Light} & \multicolumn{1}{l|}{Case-1} & No & 0.26 & 0 & 3 $\times 10^{-5}$ & 0 & 0.91 & 0 & 1.0 $\times 10^{-3}$ & 0 \\
 & \multicolumn{1}{l|}{Case-2} & No & 0.26 & 0.62 & 2.9 $\times 10^{-4}$ & 0.83 & 0.91 & 0.89 & 1.0 $\times 10^{-2}$ & 0.85 \\
 & \multicolumn{1}{l|}{Case-3} & No & 0.26 & 0.31 & 1.4 $\times 10^{-4}$ & 0.77 & 0.69 & 0.32 & 4.2 $\times 10^{-3}$ & 0.83 \\
 & \multicolumn{1}{l|}{Case-4} & No & 0.27 & 0.62 & 3$\times 10^{-5}$  & 0.16 & 0.91 & 0.89 & 1.9 $\times 10^{-3}$ & 0.33 \\ \hline
\end{tabular} \caption{Same as Table \ref{tab:results_rapid}, but for oligarchic growth.} \label{tab:results_olig}
\end{table*}

\begin{table*}[]
\begin{tabular}{|lll|llll|llll|}
\hline
\multicolumn{3}{|l|}{Pebble growth} & \multicolumn{4}{l|}{After 3 Myr or at M$_{\textrm{core}}$ = M$_{\textrm{env}}$} & \multicolumn{4}{l|}{After 10 Myr or at M$_{\textrm{core}}$ = M$_{\textrm{env}}$} \\ \hline
\multicolumn{2}{|l|}{3 au} & M$_{\textrm{core}}$ = M$_{\textrm{env}}$? & M$_{\textrm{core}}$ (M$_{\oplus}$) & x$_{\textrm{water}}$ & M$_{\textrm{env}}$ (M$_{\oplus}$) & Z$_{\textrm{env}}$ & M$_{\textrm{core}}$ (M$_{\oplus}$) & x$_{\textrm{water}}$ & M$_{\textrm{env}}$ (M$_{\oplus}$) & Z$_{\textrm{env}}$ \\ \hline
\multirow{3}{*}{Heavy} & \multicolumn{1}{l|}{Case-1} & 0.763 Myr & 12.4 & 0 & 12.4 & 0 & n.a. & n.a. & n.a. & n.a. \\
 & \multicolumn{1}{l|}{Case-2} & 0.686 Myr & 9.4 & 0.99 & 9.4 & 0.17 & n.a. & n.a. & n.a. & n.a. \\
 & \multicolumn{1}{l|}{Case-3} & 0.793 Myr & 11.6 & 0.48 & 11.6 & 0.04 & n.a. & n.a. & n.a. & n.a. \\ 
 & \multicolumn{1}{l|}{Case-4} & 0.84 Myr & 11.8 & 0.99 & 11.8  & 0.09 & n.a. & n.a. & n.a. & n.a. \\ \hline
\multirow{3}{*}{Medium} & \multicolumn{1}{l|}{Case-1} & 1.74 Myr & 15.5 & 0 & 15.5 & 0 & n.a. & n.a. & n.a. & n.a. \\
 & \multicolumn{1}{l|}{Case-2} & 1.68 Myr & 12.8 & 0.99 & 12.8 & 0.15 & n.a. & n.a. & n.a. & n.a. \\
 & \multicolumn{1}{l|}{Case-3} & 1.69 Myr & 11.9 & 0.38 & 11.9 & 0.22 & n.a. & n.a. & n.a. & n.a. \\ 
 & \multicolumn{1}{l|}{Case-4} & 1.17 Myr & 10.4 & 0.99 & 10.4  & 0.07 & n.a. & n.a. & n.a. & n.a. \\ \hline
\multirow{3}{*}{Light} & \multicolumn{1}{l|}{Case-1} & 3.47 Myr & 10.8 & 0 & 1.64 & 0 & 10.9 & 0 & 10.9 & 0 \\
 & \multicolumn{1}{l|}{Case-2} & 2.38 Myr & 7.74 & 0.99 & 7.74 & 0.55 & n.a. & n.a. & n.a. & n.a. \\
 & \multicolumn{1}{l|}{Case-3} & No & 1.54 & 0.01 & 0.01 & 0.33 & 3.18 & 0.01 & 0.35 & 0.33 \\ 
 & \multicolumn{1}{l|}{Case-4} & 3.72 Myr & 9.22 & 0.99 & 3.12  & 0.39 & 9.28 & 0.99 & 9.28 & 0.15 \\ \hline
\multicolumn{2}{|l|}{5 au} &  & M$_{\textrm{core}}$ (M$_{\oplus}$) & x$_{\textrm{water}}$ & M$_{\textrm{env}}$ (M$_{\oplus}$) & Z$_{\textrm{env}}$ & M$_{\textrm{core}}$ (M$_{\oplus}$) & x$_{\textrm{water}}$ & M$_{\textrm{env}}$ (M$_{\oplus}$) & Z$_{\textrm{env}}$ \\ \hline
\multirow{3}{*}{Heavy} & \multicolumn{1}{l|}{Case-1} & 0.91 Myr & 16.6 & 0 & 16.6 & 0 & n.a. & n.a. & n.a. & n.a. \\
 & \multicolumn{1}{l|}{Case-2} & 0.96 Myr & 14.8 & 0.99 & 14.8 & 0.14 & n.a. & n.a. & n.a. & n.a. \\
 & \multicolumn{1}{l|}{Case-3} & 0.89 Myr & 15.0 & 0.46 & 15.0 & 0.07 & n.a. & n.a. & n.a. & n.a. \\ 
 & \multicolumn{1}{l|}{Case-4} & 0.85 Myr & 15.4 & 0.99 & 15.4  & 0.1 & n.a. & n.a. & n.a. & n.a. \\ \hline
\multirow{3}{*}{Medium} & \multicolumn{1}{l|}{Case-1} & 1.74 Myr & 15.5 & 0 & 15.5 & 0 & n.a. & n.a. & n.a. & n.a. \\
 & \multicolumn{1}{l|}{Case-2} & 1.68 Myr & 12.8 & 0.99 & 12.8 & 0.15 & n.a. & n.a. & n.a. & n.a. \\
 & \multicolumn{1}{l|}{Case-3} & 1.69 Myr & 11.9 & 0.38 & 11.9 & 0.22 & n.a. & n.a. & n.a. & n.a. \\ 
 & \multicolumn{1}{l|}{Case-4} & 1.49 Myr & 10.7 & 0.99 & 10.7  & 0.20 & n.a. & n.a. & n.a. & n.a. \\ \hline
\multirow{3}{*}{Light} & \multicolumn{1}{l|}{Case-1} & No & 2.54 & 0 & 8.8 $\times10^{-3}$ & 0 & 6.35 & 0 & 0.87 & 0 \\
 & \multicolumn{1}{l|}{Case-2} & 7.69 Myr & 2.49 & 0.96 & 0.06 & 0.82 & 3.99 & 0.97 & 3.99 & 0.46 \\
 & \multicolumn{1}{l|}{Case-3} & No & 2.54 & 0.48 & 0.01 & 0.18 & 5.27 & 0.38 & 2.6 & 0.42 \\ 
 & \multicolumn{1}{l|}{Case-4} & 9.31 Myr & 2.54 & 0.96 & 1.34 $\times 10^{-2}$  & 0.33 & 5.22 & 0.98 & 5.22 & 0.23 \\ \hline
\end{tabular} \caption{Same as Table \ref{tab:results_rapid}, but for pebble growth.} \label{tab:results_pebb} 
\end{table*}

\end{appendix}

\end{document}